\documentclass[english,format=acmsmall, review=false, screen=true]{acmart}
\usepackage[T1]{fontenc}
\usepackage[latin9]{inputenc}
\pdfoutput=1
\PassOptionsToPackage{ruled}{algorithm2e}
\usepackage{color}
\usepackage{array}
\usepackage{multirow}
\usepackage{textcomp}
\usepackage{multirow}
\usepackage{algorithm2e}
\usepackage{amsbsy}
\usepackage{graphicx}
\usepackage{algorithmic}
\usepackage{setspace}
\usepackage{subfig}
\makeatletter

\setcopyright{acmcopyright}
\copyrightyear{2018}
\acmYear{2018}
\acmDOI{10.1145/1122445.1122456}

\acmJournal{JACM}
\acmVolume{37}
\acmNumber{4}
\acmArticle{111}
\acmMonth{8}
\newcommand{\redefineshortauthors}[1]{\renewcommand{\shortauthors}{#1}}

\SetAlFnt{\small}
\SetAlCapFnt{\small}
\SetAlCapNameFnt{\small}
\SetAlCapHSkip{0pt}
\IncMargin{-\parindent}

\makeatother

\usepackage{babel}

\begin{document}
% paper title
\title{}
\title[Cross-Modal Subspace Representation]{}
\title{Variational Autoencoder with CCA for Audio-Visual Cross-Modal Retrieval}

\author{Jiwei Zhang}
\orcid{0000-0003-4494-8648}
\authornote{Jiwei was involved in this work when he worked as an assistant researcher at the National Institute of Informatics, Tokyo, Japan.}
\affiliation{%
	\institution{Digital Content and Media Sciences Research Division, National Institute of Informatics}
	\streetaddress{2-1-2 Hitotsubashi}
    \city{Chiyoda-ku}
    \state{Tokyo}
    \postcode{101-8430}
    \country{Japan}}
\email{jiweizhang@nii.ac.jp}  

\author{Yi Yu}
\affiliation{%
	\institution{Digital Content and Media Sciences Research Division, National Institute of Informatics}
	\streetaddress{2-1-2 Hitotsubashi}
	\city{Chiyoda-ku}
	\state{Tokyo}
	\postcode{101-8430}
	\country{Japan}}
\email{yiyu@nii.ac.jp}

\author{Suhua Tang}
\affiliation{%
	\institution{Department of Computer and Network Engineering, Graduate School of Informatics and
Engineering, The University of Electro-Communications}
\streetaddress{1-5-1 Chofugaoka}
\city{Chofu-shi}
\state{Tokyo}
\postcode{182-8585}
\country{Japan}}
\email{shtang@uec.ac.jp}

\author{Jianming Wu}
\affiliation{%
	\institution{KDDI Research, Inc}
	\streetaddress{2-1-15 Ohara}
	\city{Fujimino}
	\state{Saitama Prefecture}
	\postcode{356-8502}
	\country{Japan}}
\email{ji-wu@kddi-research.jp}

\author{Wei Li}
\affiliation{%
	\institution{School of Computer Science, Fudan University}
	\streetaddress{220 Handan Road, Yangpu District}
	\city{Shanghai}
	\state{Shanghai}
	\postcode{356-8502}
	\country{China}}
\email{weili-fudan@fudan.edu.cn}
%   Abhishek Srivastava \\
%  MIDAS, IIIT-Delhi, India \\
%  \texttt{abhishek18124@iiitd.ac.in} \\

\begin{abstract}
Cross-modal retrieval is to utilize one modality as a query to retrieve data from another modality, which has become a popular topic in information retrieval, machine learning, and database. How to effectively measure the similarity between different modality data is the major challenge of cross-modal retrieval. Although several reasearch works have calculated the correlation between different modality data via learning a common subspace representation, the encoder's ability to extract features from multi-modal information is not satisfactory. In this paper, we present a novel variational autoencoder (VAE) architecture for audio-visual cross-modal retrieval, by learning paired audio-visual correlation embedding and category correlation embedding as constraints to reinforce the mutuality of audio-visual information. On the one hand, audio encoder and visual encoder separately encode audio data and visual data into two different latent spaces. Further, two mutual latent spaces are respectively constructed by canonical correlation analysis (CCA). On the other hand, probabilistic modeling methods is used to deal with possible noise and missing information in the data. Additionally, in this way, the cross-modal discrepancy from intra-modal and inter-modal information are simultaneously eliminated in the joint embedding subspace. We conduct extensive experiments over two benchmark datasets. The experimental outcomes exhibit that the proposed architecture is effective in learning audio-visual correlation and is appreciably better than the existing cross-modal retrieval methods.

\end{abstract}


\selectlanguage{english}%


\maketitle

\redefineshortauthors{JW. Zhang et al.}

\section{Introduction}
Many services and applications connected with IoT
involve more than one modality, which generate various data
and information in different modalities (e.g., senor, image,
video, audio, text). Consequently, these multimodal data are
accumulated over time at an unprecedented scale, which contain
useful knowledge structures for describing various events and
phenomena. This motivates us to use artificial intelligence to
model the cognitive process of human with these big-scale data,
and exploit it in cross-modal retrieval. Cross-modal representation learning is becoming more and more important for understanding the correlation among various modality data, which facilitates cross-modal retrieval between different modality data (such as text, audio, and visual). That is to say, using single modal information as a query is to retrieve the associated information of another modal. 
Searching results provided throughout several modalities can help customers to obtain comprehensive information about the desired event or topics.

Now, more and more researchers from academia and industry are starting to pay interest to cross-modal retrieval research.
The major challenge of the cross-modal retrieval task is how to measure the content similarity between different modality data in the joint space, which is known as the heterogeneity gap.
%The approach of representation learning is usually used to bridge the heterogeneity gap between special modalities. It aims to project data samples from different modalities into a common representation subspace, where loss functions can be used to measure their similarity directly.
Many efforts have been carried out recently for cross-modal retrieval \cite{andrew2013deep,wang2015deep,wang2016comprehensive,wei2016cross,xu2019deep,yu2012automatic,yu2019deep,peng2017overview,zhang2018unsupervised,wang2015joint,zhang2017generalized,wu2017joint} between different modalities such as text-image \cite{wang2016comprehensive}, video-text \cite{xu2019deep}, and audio-lyrics \cite{yu2019deep} with different levels of semantics. 
%Cross-modal retrieval methods between audio and video have been studied in previous works such as \cite{yu2012automatic,shah2014advisor} where a video can become much more attractive by adding a matching soundtrack to it. 

%A variety of cross-modal retrieval methods \cite{peng2017overview,zhang2018unsupervised,wang2015joint,zhang2017generalized,wu2017joint} have been developed, and their purpose is to find a common representation subspace via different methods. 
Traditionally, linear projections are generally computed for measuring the correlation between different modalities. . For example, Canonical Correlation Analysis (CCA) \cite{hotelling1992relations,hardoon2004canonical,leurgans1993canonical} is the most well-known learning methods. It finds the linear transformation of the two modalities of the input data by maximizing the pairwise correlation in the common subspace. 

%However, it is impossible to extract useful features only by applying linear because of the non-linear relationship of multimedia data in the real world. Therefore, the kernel-based approaches \cite{akaho2006kernel,wang2015large} are used to solve this issue, but how to select a suitable kernel function is still difficult for researchers.

Benefiting from the speedy improvement of deep neural networks (DNN) technology, a lot of researchers have utilized DNN to capture nonlinear correlation and guide the model to learn the common representation subspace of multimodal data. Deep Canonical Correlation Analysis (DCCA) \cite{andrew2013deep}, is extensively utilized to learn complicated nonlinear transformations of of different modalities. Some supervised deep cross-modal learning networks based on semantic category information are proposed to guide the model to learn extra discriminative representations. It aims to enable high-level semantic separation between different semantics in the common representation subspace. C-DCCA \cite{yu2018category,zeng2018audio} utilizes a deep learning network to extend the standard CCA, learning the nonlinear projection between pairs of different modality data while effectively preserving semantic information.
%In the above-mentioned existing studies: 
%In \cite{wang2016effective}, Wang et al. propose a Multi-modal Deep Neural Network (MDNN) based on a deep Convolutional Neural Network (CNN) and a Neural Language Model (NLM) to learn mapping functions for the image modality and the audio modality, respectively. 

%\begin{figure}[!h]
%\centering 
%\includegraphics[width=.8\textwidth]{images/cross-modal-correlation-learning}
%\caption{Illustrations of mainstream framework for the audio-visual representation learning in the common subspace, which aims to project the heterogeneous data of different modalities from their own feature spaces into one common subspace, where similarity measurement can be directly adopted to establish correlation on the cross-modal data, and the learned features are fused and utilized for the calculation of cross-modal similarity.}
%\label{fig:representation}
%\end{figure}

%1) Various schemes are used to learn linear or non-linear projections; 

%2) Most of the semantic information was used to learn the intra-modal semantics of audio and visual samples; 

%3) Euclidean distance is used to measure the difference between the representations of audio-visual pairs to guide cross-modal learning; 

%4) Maximize the pairwise correlation in the common subspace of two modality data.

The success of present state-of-the-art methods has strongly relied on a giant quantity of high-quality labeled data. However, due to the wide range of information sources, it causes many noises and a lack of information. If machine learning methods ignore these uncertainties, it will inevitably lead to many meaningless statistical results. Therefore, it is necessary to use probabilistic modeling methods to deal with the uncertainty in the data. 
%To make the decoder robust to noise, we have introduced the idea of VAE \cite{kingma2013auto}. In addition, we use the VAE network to guarantee that the features extracted by the deep network encoder can still reflect the structural distribution of the original data.
In MS-VAE \cite{zhu2021learning}, Zhu et al. presented a self-supervised architecture based on the VAE network (without semantic category information) to learn audio-visual correlation to realize cross-modal retrieval tasks.

%In these previous cross-modal methods, semantic information was used to learn the intra-modal semantics of audio and visual samples. Euclidean distance is used to measure the difference between the representations of audio-visual pairs to guide cross-modal learning. 
%It is notable that even though the classification information has been used in these approaches, the classification information is only used to learn discriminative features within each modality or between inter modalities. The semantic information is not fully exploited in these cross-modal learning approaches. Moreover, they cannot well capture the effective features of cross-modal retrieval.

Currently, existing methods have used semantic information to find out discriminative features from intra-modal or inter-modal data. However, in these cross-modal correlation learning methods, semantic category information has not been effectively utilized. Therefore, cross-modal representation learning still faces the following challenges:

1) How to effectively use the correlation between inter-modal and intra-modal data to learn the shared subspace representation better to enhance retrieval performance.

2) How to guarantee that the features encoded by the deep network can still reflect the structural distribution of the original data.

3) How to deal with the noise and missing data that may exist in the data, as the deep neural network ignores the uncertainty in the data.

To challenge these issues, in this paper, we propose a novel variational autoencoder (VAE) architecture for simultaneously mitigating the cross-modal discrepancy from intra-modal and inter-modal data. Probabilistic modeling of the distribution of input data is to generate latent variables that observe Gaussian unit distribution. In addition, it ensures that the feature extraction encoder can still well reflect the features of the original data. To obtain this goal, it minimizes the discriminative loss of samples in the semantic category subspace so as to supervise our architecture to learn discriminative features. In addition, the decoder is exploited to reconstruct latent variables. Therefore, each pairwise semantic information and the category information are entirely exploited to make sure that the learned representation is both discriminative in semantic structure and invariant across modalities.

%Furthermore, it simultaneously minimizes the modality invariance loss and uses a weight-sharing strategy to learn modality invariant features in the common representation subspace. Following this learning strategy, both the pairwise semantic information and the category information are fully exploited to ensure that the learned representation is both discriminative in semantic structure and invariant across modalities.

Our proposed cross-modal retrieval architecture between audio and video has several significant contributions, as follows:

i) The idea of VAE, while preserving the semantic structure of the original data, can also effectively learn the audio-visual correlation in the common representation subspace. Probabilistic modeling method is applied to avoid the inability to handle the possible noise and lacking information in the audio-visual data.

ii) Audio-visual pairwise-level and category-level mutual latent spaces are separately constructed by CCA, which can enhance audio-visual correlation embedding and mitigate the cross-modal discrepancy between audio-visual data.

iii) Several loss constraints are proposed to optimize the audio-visual representation learning in the mutual latent spaces, which can reinforce both audio-visual pairwise-level and category-level correlation. 

iv) Extensive experiments on benchmark datasets present that our approach is superior to the current state-of-the-art audio-visual cross-modal retrieval methods.
\section{Related works}
Since different modalities usually have inconsistent distributions and feature representations, it is essential to learn a common space to bridge heterogeneous data and measure their correlation. The purpose is to project each modality's feature into a common representation subspace that can directly evaluate the correlation between different modalities. Various strategies have been proposed to analyze this common representation subspace. In this paper, we focus on real-valued representation learning to learn the common space to achieve audio-visual cross-modal retrieval. This category includes unsupervised approaches, semi-supervised  approaches, and supervised approaches. 

The unsupervised methods only use co-occurrence information to learn common representations for different modality data, such as CCA \cite{hotelling1992relations}, has been widely utilized in cross-modal retrieval tasks via associating features of different modalities in a common representation space. Rasiwasia et al \cite{rasiwasia2010new} proposed to maximize the correlation between visual features and text features in the common representation space via utilizing the CCA method. The methods of correspondence autoencoder (Corr-AE) \cite{feng2014cross} and deep canonically correlated autoencoder (DCCAE) \cite{wang2015deep} are representative works of this subcategory. In \cite{zhu2021learning}, the authors proposed  to give visual or audio modality as input data to generate a pair of corresponding audio-visual data. The pair-based method is to learn the correlation between samples via utilizing paired different modalities data. The authors in \cite{zhai2012multiview} suggest multi-view metric learning (MVML-GL) methods with global consistency and local smoothness.
%The methods of CCA\cite{hotelling1992relations}, Deep CCA (DCCA)\cite{andrew2013deep} , Correspondence Auto-encoder (Corr-AE)\cite{feng2014cross} and Deep Canonically Correlated Auto-encoder (DCCAE)\cite{wang2015deep} are representative ones of this subclass. 

Semi-supervised methods can exploit useful unlabeled data to improve supervised learning tasks when the labeled data are scarce or expensive. In Adaptively Unified Semi-supervised Learning (AUSL) for cross-modal retrieval in \cite{zhang2017adaptively}, the authors present a semi-supervised approach which uses both relevance of class labeled and unlabeled data to strengthen multi-modal correlation. In adaptive semi-supervised feature selection (ASFS) for cross-modal retrieval in \cite{yu2018adaptive}, the authors learns potential information from unlabeled data to analyze the correlation between different modality data. In semi-supervised cross-modal retrieval (SSCMR) with label prediction in \cite{mandal2019semi}, the author applied the category prediction layer to guide the model to learn the common representation of different modalities, so as to achieve cross-modal retrieval tasks.

To learn more discriminative common representations, supervised methods exploit semantic category information to distinguish the samples from different categories. Supervised methods can be used to calculate the distance between samples of different categories, and the distance between samples of the same category should be as close as possible. In deep supervised cross-modal retrieval (DSCMR) \cite{zhen2019deep}, an approach of using semantic category information to  learning discriminative features is proposed. In addition, to preserve cross-modal similarity, many existing cross-modal retrieval methods \cite{wang2016effective} guide the model to reduce the distance of each pair of modal data in the common Hamming space. In adversarial cross-modal retrieval (ACMR) \cite{wang2017adversarial}, the category information is utilized to learn the discriminative features during the feature projection process.

This paper aims to fully apply both pairwise-level and category-level audio-visual information as constraints to guide the model to learn more discriminative and modal-invariant representations of different modalities data. Our proposed architecture bridges the heterogeneity gap while capturing discriminative features and ensuring that the representation extracted by the encoder still reflects the feature of the original data to enhance the accuracy of cross-modal retrieval. In particular, paired audio-visual and category-level information are exploited to simultaneously guide the model to learn more discriminative features. Probabilistic modelling methods are exploited to deal with the uncertainty in the data and make the decoder robust to noise. In addition, CCA is suggested to learn audio-visual correlation embedding in the mutual latent space. the idea of VAE is used to guarantee that the features extracted by the deep encoder can nevertheless reflect the structural distribution of the original data.
\section{Our Method}
We first explain the motivation of cross-modal subspace representation based on VAE with discriminative loss function. 
Then we introduce the proposed architecture. Finally, we analyze the objective loss function and the detailed training process.

\begin{figure*}[!h]
	\centering 
	\includegraphics[width=1.0\textwidth,height=0.30\textwidth]{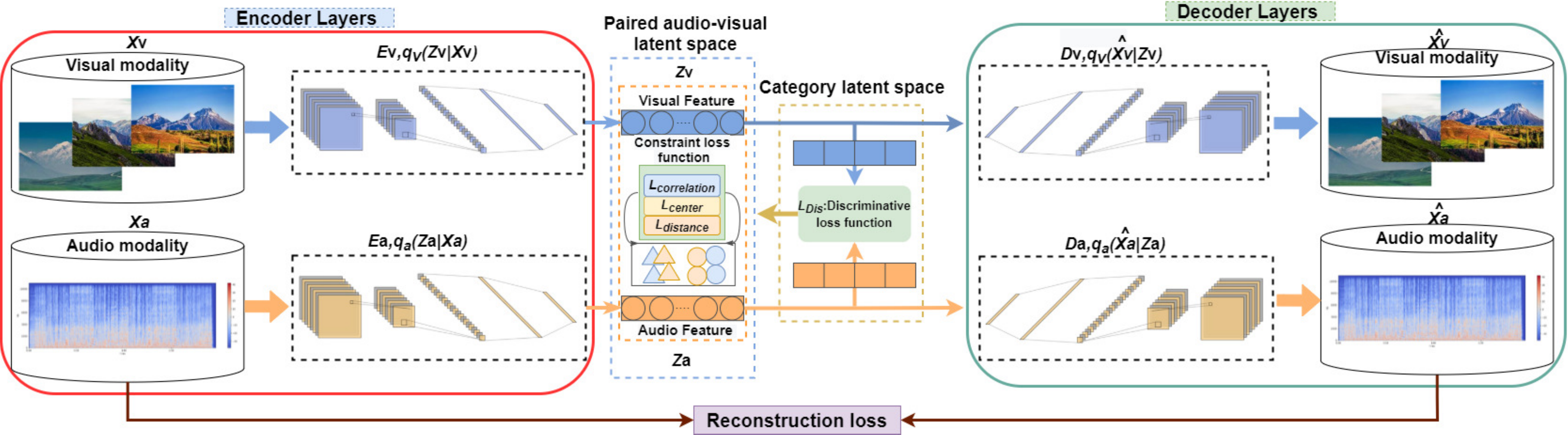}
	\caption{The overall architecture of the proposed method. $\mathbf{X}_{v}$ and $\mathbf{X}_{a}$ are the input data from two modalities. $\widehat{\mathbf{X}}_{v}$ and $\widehat{\mathbf{X}}_{a}$ are the reconstruct data. $\mathbf{E}_{v}$ and $\mathbf{E}_{a}$ are deep convolutional encoders. $\mathbf{Z}_{v}$ and $\mathbf{Z}_{a}$ are latent representations from the outputs
		of $\mathbf{E}_{v}$ and $\mathbf{E}_{a}$. $\mathbf{D}_{v}$ and $\mathbf{D}_{a}$ are decoders.}
	\label{fig:proposedmethod}
\end{figure*}
\subsection{Motivations}
There are various cross-modal retrieval approaches based on CCA, such as
\cite{jin2015cross,zhang2007cross}. However, these methods can only learn linear features. With the development of deep neural networks, Deep-CCA and its variants such as C-DCCA \cite{yu2018category,zeng2018audio} and TNN-C-CCA \cite{zeng2020deep} are proposed to solve the non-linear correlation learning. However, in these existing methods, it is not considered whether the features extracted by the neural network encoder still retain the general structure of the original data. Therefore, in this work, the decoder is utilized in the architecture to reconstruct latent features from the deep convolutional encoder, which ensures that the underlying features can reflect the structure of the original data to improve retrieval performance. In addition, the deep neural network ignores the uncertainty in the data and cannot deal with the noise and missing information that may exist in the data. To make the decoder robust to noise, Gaussian noise is added to the output of the encoder and probabilistic modeling methods are used to deal with the uncertainty in the data.
\subsection{VAE-CCA Architecture}

The proposed VAE with CCA architecture mainly consists of three parts as shown in Fig. \ref{fig:proposedmethod}: two branch encoder layers, mutual latent space, and two branch decoder layers.

\subsubsection{Encoder layers} In our VAE-CCA model, we assume that two modality data are $\mathbf{X}_{v}=$ $\left\{x_{v}^{i}\right\}_{i=1}^{m} \in \mathbb{R}^{d_{v} \times m}, \mathbf{X}_{a}=\left\{x_{a}^{i}\right\}_{i=1}^{m} \in \mathbb{R}^{d_{a} \times m}$, where $m$ is the number of samples. ${d}_{v}$ and ${d}_{a}$ are the corresponding dimensions of the visual modality and the audio modality. We set deep convolutional encoders to three layers for two modaities. $\mathbf{X}_{v}$ and $\mathbf{X}_{a}$ are fed as inputs to deep convolutional encoders to process and we can obtain the latent representations $\left(\mathbf{Z}_{v} \mid \theta_{e_{v}}\right) \in \mathbb{R}^{o \times m}$ and $\left(\mathbf{Z}_{a} \mid \theta_{e_{a}}\right) \in \mathbb{R}^{o \times m}$, where $\theta_{e_{v}}$, $\theta_{e_{a}}$ are respectively the parameters of deep convolutional encoders of visual and audio, and $o$ is the output dimension of deep convolutional encoders. 
\subsubsection{Mutual latent space} Data from different modalities have different statistical characteristics and feature representations. Therefore, for cross-modal retrieval tasks, they cannot directly measure the correlation between different modalities data. In our work, we project the data of two modalities into common subspace, where we calculate the correlations between $\mathbf{Z}_{v}$ and $\mathbf{Z}_{a}$ with the following expression:

$$
\underset{\theta_{e_{v}}, \theta_{e_{a}}}{\arg \max } \operatorname{corr}\left(\mathbf{Z}_{a}, \mathbf{Z}_{v}\right)=\underset{\theta_{e_{a}}, \theta_{e_{v}}}{\arg \max } \frac{\operatorname{cov}\left(\mathbf{Z}_{a}, \mathbf{Z}_{v}\right)}{\sqrt{var\left(\mathbf{Z}_{a}\right)} \sqrt{var\left(\mathbf{Z}_{v}\right)}}
$$
where $\operatorname{corr}(\cdot)$ is the correlation between $\mathbf{Z}_{a}$ and $\mathbf{Z}_{v}$. $\operatorname{cov}\left(\mathbf{Z}_{a}, \mathbf{Z}_{v}\right)$ is the covariance of $\mathbf{Z}_{a}$ and $\mathbf{Z}_{v}$, and $var\left(\mathbf{Z}_{i}\right)$ is the variance of $\mathbf{Z}_{i}, i=a,v.$

\subsubsection{Category latent space}The label matrix is denoted by $\mathbf{Y_{i}}=\left[y_{i 1 }, y_{i 2 }, \ldots, y_{i c }\right] \in \mathbb{R}^{c}$, where $c$ is the total number of categories. If the $i$th sample belongs to the $j$th category, $y_{i j }$ = 1, otherwise $y_{i j }$ = 0, we convert the category into one-hot form.

\subsubsection{Decoder layers} The decoder of each modality is composed of three layers of neural networks, which aims to reconstruct the latent representations from a shared wight layer and maintain the structural characteristics of the original data. We can obtain the outputs 
$\left(\widehat{\mathbf{X}}_{v} \mid \theta_{d_{v}}\right)$ and $\left(\widehat{\mathbf{X}}_{a} \mid \theta_{d_{a}}\right)$, where $\theta_{d_{v}}$ and $\theta_{d_{a}}$are network parameters of the decoders.

\subsection{Objective Loss Function Analysis}
The goal of our architecture is to achieve the latent representation that can learn audio-visual correlation in the mutual space to mitigate the discrepancy between audio and visual data in different feature spaces. To do this, multiple loss functions are proposed to optimize our architecture.

\subsubsection{VAE model Loss \cite{tian2018audio}}
The objective function of our proposed architecture in reconstructing the data part is similar to the original VAE, where the log-probability $\log p\left(X\right)$ of the reconstruction data pair $\mathbf{X}_{i}$ is maximized from the desired mutual latent space $\mathbf{Z}_{i}$, and $i$ represents either the modality $a$ (audio) or $v$ (visual). We can get the variational lower bound of the objective function based on the data reconstruction of the original VAE network as follows:

\begin{equation}
	\log p(X) \geq \mathbb{E}_{Z_{i} \sim q_{i}\left(Z_{i} \mid X_{i}\right)}\left[\log p\left(X \mid Z_{i}\right)\right]-\operatorname{KL}\left(q_{i}\left(Z_{i} \mid X_{i}\right) \| p\left(Z_{i}\right)\right)
\end{equation}

During the training process, the audio-visual pair from audio or visual input is reconstructed at each epoch. $\mu_{i}$ and $\sigma_{i}$ for the Gaussian distribution are returned by the encoder, and $\mathcal{Z}_{i}$ is sampled from $\mathcal{N}\left(\mu_{i}, \sigma_{i}\right), i=a,v$.
In order to ensure the validity of the representation extracted by the encoder layer, we further add the decoders to separately reconstruct audio and visual data. The representations $\mathbf{Z}_{v}$ and $\mathbf{Z}_{a}$ from the shared weight layer are fed to the decoders and we can obtain the reconstruct data $\left(\widehat{\mathbf{X}}_{v} \mid \theta_{d_{v}}\right)$ and $\left(\widehat{\mathbf{X}}_{a} \mid \theta_{d_{a}}\right)$. Minimizing errors between $\widehat{\mathbf{X}}_{a}$, $\widehat{\mathbf{X}}_{v}$ reconstructed data and 
$\mathbf{X}_{a}$, $\mathbf{X}_{v}$ original data can optimize the VAE networks. 
Therefore, the reconstruction loss for the network is:
\begin{equation}
	\mathcal{L}_{rec}=\min _{\theta_{d_{v}}, \theta_{d_{a}}} \sum_{i=v,a}\left\|\mathbf{X}_{i}-\widehat{\mathbf{X}}_{i}\right\|_{\mathrm{F}}^{2}
\end{equation}
We adopt the mean square error (MSE) \cite{wang2009mean}  algorithm as the reconstruction loss function to calculate the audio-visual pair deviation between the reconstructed data $\widehat{\mathbf{X}}_{i}$ and the original data $\mathbf{X}_{i}$. In the stage of training the VAE network to reconstruct the data, the total loss includes reconstruction loss and KL divergence.
\begin{equation}
	\mathcal{L}_{\text {V}}= \mathcal{L}_{\text {rec}}+ \mathcal{L}_{\mathrm{KL}}
\end{equation}
KL in the formula is used to represent the Kullback-Leibler divergence function, which is defined as: $\mathrm{KL}(p(x) \| q(x))=\int_{x} p(x) \log \frac{p(x)}{q(x)}$ is utilized to calculate the similarity between the distribution of audio-visual data.
\subsubsection{$\mathcal{L}_{corr}$-correlation Loss \cite{mitchell1992spatial}}
In addition, we directly calculated the correlation of all paired audio-visual samples from the two modalities in the paired audio-visual latent space. As for the audio-visual pair data, we calculate the inter-modality and intra-modality correlation in the common representation subspace  and then maximize the log-likelihood of the correlation:
$$\underset{\theta_{e_{v}}, \theta_{e_{a}}}{\arg \max } \operatorname{inter-modality}_{discrimination}\left(\mathbf{Z}_{a}, \mathbf{Z}_{v}\right)=$$
$$\frac{1}{n^{2}} \sum_{a, v}^{n}\left(\log \left(1+e^{\tau_{a v}}\right)-\kappa_{a v} \tau_{a v}\right)$$

where $\tau_{a v}=\frac{1}{2} corr\left(\mathbf{Z}_{a}, \mathbf{Z}_{v}\right)$, $\kappa_{a v}=1\left\{\mathbf{P}_{a}, \mathbf{P}_{v}\right\}$.

$$\underset{\theta_{e_{v}}, \theta_{e_{v}}}{\arg \max } \operatorname{intra-modality}_{discrimination}\left(\mathbf{Z}_{v}, \mathbf{Z}_{v}\right)=$$
$$\frac{1}{n^{2}} \sum_{v, v}^{n}\left(\log \left(1+e^{\gamma_{v v}}\right)-\zeta_{v v} \gamma_{v v}\right)$$

where $\gamma_{v v}=\frac{1}{2} corr\left(\mathbf{Z}_{v}, \mathbf{Z}_{v}\right)$, $ \zeta_{v v}=1\left\{\mathbf{P}_{v}, \mathbf{P}_{v}\right\}$.

$$\underset{\theta_{e_{a}}, \theta_{e_{a}}}{\arg \max } \operatorname{intra-modality}_{discrimination}\left(\mathbf{Z}_{a}, \mathbf{Z}_{a}\right)=$$
$$\frac{1}{n^{2}} \sum_{a, a}^{n} \left(\log \left(1+e^{\psi_{a a}}\right)-\phi_{a a} \psi_{a a}\right)$$

where $\psi_{a a}= \frac{1}{2} corr\left(\mathbf{Z}_{a}, \mathbf{Z}_{a}\right)$, $ \phi_{a a}=1\left\{\mathbf{P}_{a}, \mathbf{P}_{a}\right\}$.
Generally speaking, our total cross-modal correlation loss function is defined as:
\begin{equation}
	\begin{aligned}
		\mathcal{L}_{corr} =  &\underset{\theta_{e_{v}}, \theta_{e_{a}}}{\arg \max } \operatorname{inter-modality}_{discrimination}\left(\mathbf{Z}_{a}, \mathbf{Z}_{v}\right)\\
		&+\underset{\theta_{e_{v}}, \theta_{e_{v}}}{\arg \max } \operatorname{intra-modality}_{discrimination}\left(\mathbf{Z}_{v}, \mathbf{Z}_{v}\right)\\
		&+\underset{\theta_{e_{a}}, \theta_{e_{a}}}{\arg \max } \operatorname{intra-modality}_{discrimination}\left(\mathbf{Z}_{a}, \mathbf{Z}_{a}\right)
	\end{aligned}
\end{equation}
$corr(\cdot)$ is utilized to calculate the similarity between audio-visual modality data, and $1\{\cdot\}$ is an indicator function. If the audio-visual modality data from the same category, its value is 1, otherwise it is 0.

\subsubsection{$\mathcal{L}_{dist}$-distance Loss}
We adopt minimizing the distance between all audio-visual pairs in the latent subspace representation to reduce the cross-modal heterogeneity difference between audio-visual modalities. Specifically, the Frobenius norm is used to directly measure the distance of all sample pairs in the paired audio-visual latent subspace. Finally, we define the modal invariance loss formula as follows:
\begin{equation}
	\mathcal{L}_{dist}=\frac{1}{n}\|\mathbf{{Z}_{v}}-\mathbf{{Z}_{a}}\|_{F} 
\end{equation}
\subsubsection{$\mathcal{L}_{discr}$-discriminative Loss \cite{zhen2019deep}}
To maintain the distinction of audio-visual samples from different categories after the features are mapped to the common space, we apply a simple linear layer classifier to predict the category of audio-visual modality samples projected in the latent space. Specifically, we connect a linear layer after the audio and visaul encoder networks. The classifier makes use of the representation of the training data in the common subspace as input and generates a $c$-dimensional  predictive category for each audio-visual sample. In the latent subspace of the semantic category, we adopt the Frobenius norm function to calculate the discriminative loss.
\begin{equation}
	\mathcal{L}_{discr}=\frac{1}{n}\left\|\mathbf{P_{a}}-\mathbf{y_{a i}}\right\|_{F}+\frac{1}{n}\left\|\mathbf{P_{v}}-\mathbf{y_{v i}}\right\|_{F}
\end{equation}
where $\|\cdot\|_{F}$ refers to the Frobenius norm, $P_{a}$ and $P_{v}$ are the projection matrix of the linear classifier, and $n$ represents the batch size.

\subsubsection{Center Loss \cite{wen2016discriminative}}
In order to improve the discriminative power of deep learning features,
we use an effective central loss function, which keeps the features of different categories separable while minimizing intra-category variations, as formulated in Equation 7.
\begin{equation}
	\mathcal{L}_{center}=\frac{1}{2} \sum_{i=1}^{m}\left\|\boldsymbol{Z}_{v}-\boldsymbol{c}_{y_{i}}\right\|_{2}^{2} + \frac{1}{2} \sum_{i=1}^{m}\left\|\boldsymbol{Z}_{a}-\boldsymbol{c}_{y_{i}}\right\|_{2}^{2}
\end{equation}
The $\boldsymbol{c}_{y_{i}} \in \mathbb{R}^{d}$ refers to the $y_{i}$th  category of the feature. The formulation
effectively characterizes the intra-category variations. 

According to analysis of the loss function for each part, the final objective function can be summarized as follows.
\begin{equation}
	\mathcal{L}_{\text {total }}=\mathcal{L}_{\text {discr}}+\lambda_{1}\mathcal{L}_{\text {V}} + \lambda_{2} \mathcal{L}_{\text {corr}}+\lambda_{3} \mathcal{L}_{\mathrm{dist}}+\lambda_{4} \mathcal{L}_{\text {center}}
\end{equation}
In our experiments, according to the experimental results of parameters analysis, we set  $\lambda_{1}= 0.0001$, $\lambda_{2}=0.001$ and $\lambda_{3}=0.1$, and $\lambda_{4}=0.01$. We apply the gradient descent algorithm to optimize the objective loss function of the proposed architecture.

\section{Training strategy}
In the proposed architecture, we utilize two steps to train the architecture and optimize the whole network parameters.

\subsection{Pre-training VAE Network}
In the first step, we pre-train the VAE network utilizing Equation 3. We feed the cross-modal data ${X}_{a}$ and ${X}_{v}$ to deep convolutional encoders ${E}_{a}$ and ${E}_{v}$, and acquire the reconstruction data $\widehat{{X}}_{a}$ and $\widehat{{X}}_{v}$ from the decoders ${D}_{a}$ and ${D}_{v}$. 

In the second step, We set the value of the learning rate to $3.5 \times 10^{-4}$. We adopt MSE to modify the objective function, where the error between the original data and the reconstructed data is minimized, the overall loss function is used to optimize the network and update encoders parameters $\theta_{e_{a}}$, $\theta_{e_{v}}$ and decoders parameters $\theta_{d_{a}}$, $\theta_{d_{v}}$. The smaller the value of MSE is, the better the prediction performance of the proposed architecture is.

\subsection{Training Entire Network}
Finally, we train the entire network using Equation 8, minimizing the total loss including the discriminative loss, crrelation loss,distance loss, center loss and the VAE network loss to update model parameters $\theta_{e_{a}}$ and $\theta_{e_{v}}$ , 
$\theta_{d_{a}}$ and $\theta_{d_{v}}$. 
Algorithm 1 is the training process of our entire architecture.

\section{Experiments}
In order to evaluate the performance of our proposed architecture, we conduct the experiment by comparing with ten remarkable baseline approaches on VEGAS and AVE datasets.
Specifically, we first describe two datasets used in our paper, followed by the evaluation results. 

\begin{algorithm}
	\caption{The discriminative feature learning algorithm }
	\setstretch{1.35}
	\SetAlgoLined
	\textbf{Input:} Visual samples for current batch: $\mathbf{X}_{v}=$ $\left\{x_{v}^{1},...,x_{v}^{n}\right\}$;\\ Audio samples for current batch, $\mathbf{X}_{a}=$ $\left\{x_{a}^{1},...,x_{a}^{n}\right\}$;\\Correseponding labels for current batch,$\mathbf{Y}=$ $\left\{y_{1},...,y_{n}\right\}$;\\ 
	\textbf{Output:} The optimised parameters $\theta$. \\ 
	\textbf{Initialize:} $\lambda_{1}$, $\lambda_{2}$, $\lambda_{3}$ and $\lambda_{4}$; learning rate = $3.5 \times 10^{-4}$; the number of iteration t =0.\\    
	\While{not converge}{
		$t \gets t+1$ \\ 
		Train the VAE network using Equation $3$ \\   
		Optimize network parameters $\theta_{e_{v}}$ and  $\theta_{e_{a}}$ of encoders and $\theta_{d_{v}}$ and $\theta_{d_{a}}$ of decoders \\ 
		Renew the parameters of the sub-networks, by mininsing Equation $8$. \\ 
	}
	
\end{algorithm}

\subsection{Experiment Setup}
\subsubsection{Datasets Settings}
Dataset with the audio-visual pairwise correlation and semantic category information are desired in the experiment.  
Therefore, we use the VEGAS \cite{zhou2018visual} dataset and the Audio-Visual Event (AVE) \cite{tian2018audio} dataset.

The Audiovisual Events (AVE) data set is a subset of AudioSet \cite{gemmeke2017audio}, consisting of 28 event categories and a total of 4,143 videos. Each video has a period of 10s and utilizes audio-visual events as marker boundaries. The videos in the AVE dataset contain at least one audio-visual event with a duration of two seconds. The AVE dataset consists of various audiovisual event sounds in various fields such as human activities, animal activities, and music performances (for example, women talking, dog barking, playing guitar, etc.). In our experiment, we remove the music-related categories and keep 15 categories (Clock,motorcycle, train horn, bark, cat, bus, rodents/rats, toilet flush, acoustic guitar, frying, chainsaw, horse, helicopter, infant
cry, truck)

The VEGAS dataset is a subset of the Google Audioset \cite{gemmeke2017audio} with Amazon Mechanical Turk clean data, including 10 natural sound categories (chainsaw, helicopter, drum, printers, fireworks, dog, and etc.). The video duration is between 2-10 seconds, with an average of 7 second. It contains 28,103 videos and each video is marked by a single label. In our experiments, we used 5,621 videos for testing and 22,482 videos for training to evaluate our architecture. Finally, we summarize the statistical results of the two datasets in Table \ref{tab:dataset}.

\begin{table}[!ht]
	\caption{Two benchmark data sets were used in our experiment, where $N_{train}$ and $N_{test}$ represent the number of training and testing audio-visual pairs, respectively. $c$ represents the number of categories, $D_{audio}$ and $D_{visual}$ are the dimensions of the audio-visual features extracted by the VGGNet and VGGish networks, respectively.}
	\centering
	% \resizebox{\linewidth}{!}
	\begin{tabular}{c|c c c c c }\hline
		Dataset        & $N_{train}$   & $N_{test}$    & c   & $D_{visual}$   & $D_{audio}$   \\ \hline
		VEGAS          & 22482    & 5621     & 10  & 1024      & 128  \\ 
		AVE            & 1766     & 189      & 15  & 1024      & 128   \\ \hline
	\end{tabular} %}

\label{tab:dataset}
\end{table}
\begin{figure}[tbp]
\centering
\subfloat[]{\label{fig:a}\includegraphics[width = 0.5\textwidth]{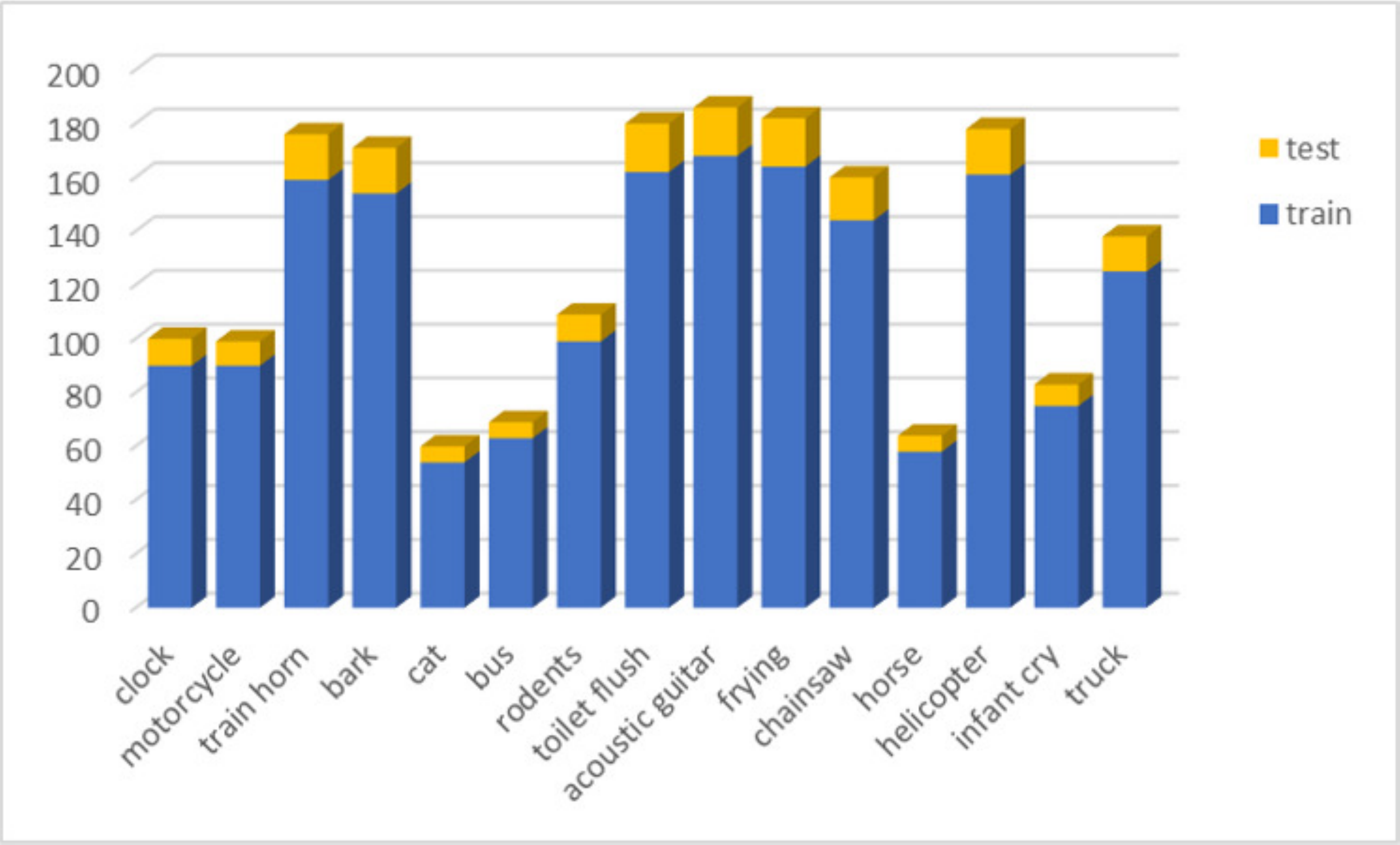}}\quad
\subfloat[]{\label{fig:b}\includegraphics[width = 0.5\textwidth]{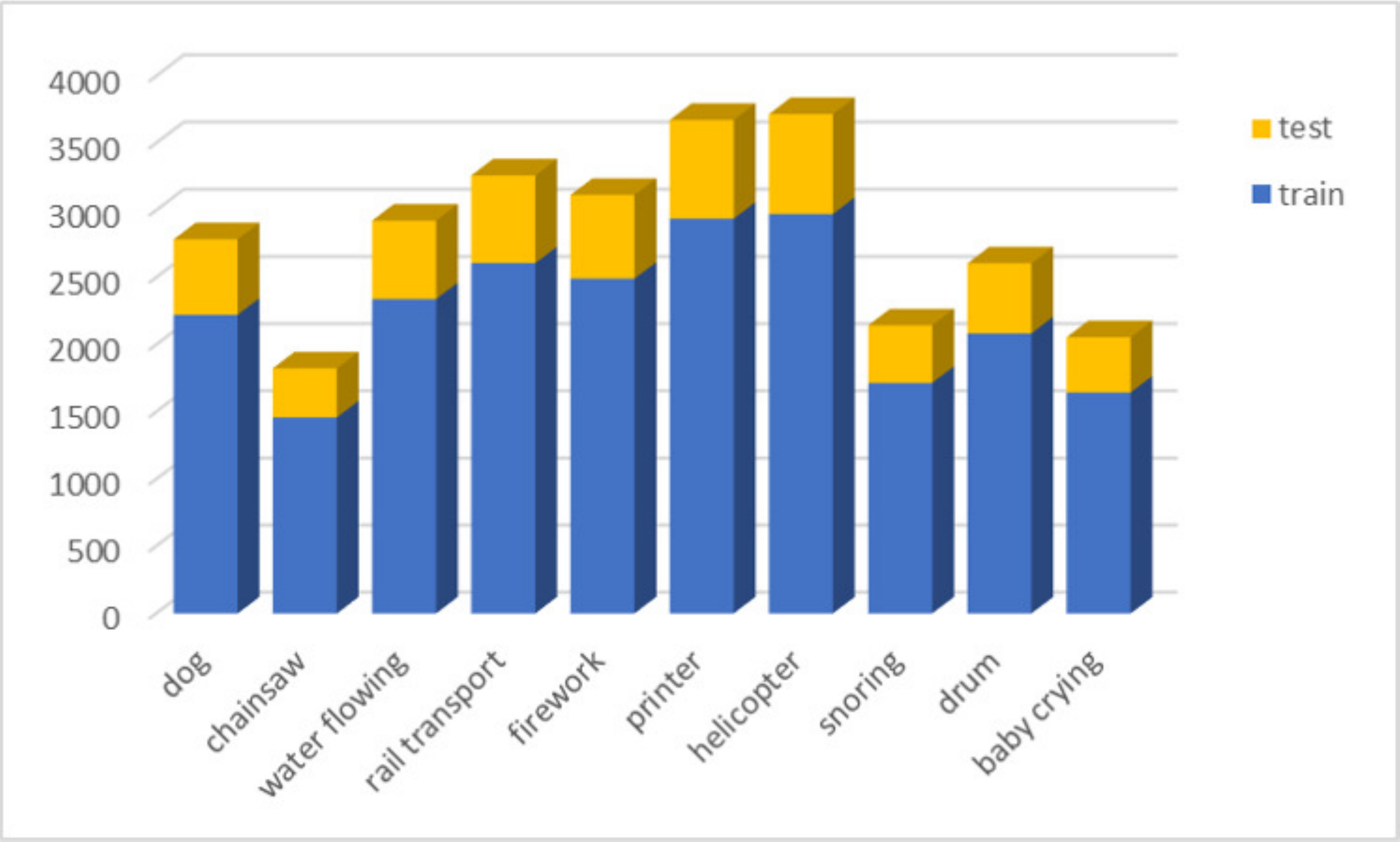}}	
\caption{ The number of videos with different categories in training set and testing set in (a) AVE dataset, and (b) VEGAS dataset.}
\end{figure}
\subsubsection{Implementation details}
In this work, our proposal architecture has two branch VAE networks, one of which is used for visual modality and the other is used for audio modality. The specific configuration of our proposed method is shown in Table \ref{tab:Configuration}. First, we utilize 19-layer VGGNet \cite{simonyan2014very} to extract visual features of 512 dimensions, and a VGGish \cite{hershey2017cnn} to extract audio features of 128 dimensions. Then an encoder is applied in each sub-network to project different modality samples into a common subspace to learn the correlation of different modalities. A linear classifier layer is connected after each encoder to guide the network to obtain extra discriminative features. Finally, we reconstruct the learned latent representation into the original data dimension through the decoder. We apply GTX 2080 Ti GPU to train the proposed VAE network, and utilize the ADAM \cite{kingma2014adam} to optimize the network parameter, the preliminary learning rate is set to $3.5 \times 10^{-5}$, and the maximum number of training times is set to five hundred.

\begin{table*}[!ht]
\centering
\caption{Configuration of our proposed method.}
% \resizebox{\linewidth}{!}
\begin{tabular}{c|c|c }\hline
	&Visual Branch  &Audio Branch \\ \hline
	Input        & 1024-D    & 128-D\\  \hline
	Layer 1 (Encoder)      & 512, fully-connected, linear    & 512, fully-connected, linear\\  \hline
	Layer 2 (Common subspace layer)     & 64, fully-connected, linear     & 64, fully-connected, linear\\ \hline
	Layer 3 (Semantic category layer)     & 10, fully-connected, linear      & 10, fully-connected, linear\\  \hline
	Layer 4 (Decoder)    & 64, fully-connected, linear      & 64, fully-connected, linear  \\  \hline 
	Layer 5 (Decoder)    & 512, fully-connected, linear    & 512, fully-connected, linear\\ \hline
	Layer 6 (Decoder)    & 1024, fully-connected, linear     & 128, fully-connected, linear \\ \hline
\end{tabular} %}
\label{tab:Configuration}
\end{table*}
\subsubsection{Warmup Learning}
The performance of the deep learning model is closely related to the setting of the learning rate. The proposal framework initially conducts model training with a large and constant learning rate. In the actual model training, first, we linearly increase the learning rate from $3.5 \times 10^{-5}$ to $3.5 \times 10^{-4}$ in the first 10 periods. Then, the learning rate decays to $3.5 \times 10^{-5}$ in the 40th period, and is set to $3.5 \times 10^{-6}$ in the 70th period and later.

\subsection{Experimental Results}
\subsubsection{Evaluation Metric}
We utilize the cosine value on the VEGAS and AVE datasets to calculate the retrieval similarity between different modalities, and adopt the average of all returned accuracy (mAP) as the evaluation metric. mAP is a performance evaluation standard widely used in cross-modal retrieval research \cite{peng2016cross,wang2017adversarial,wang2016comprehensive}. It measures the ranking information and accuracy in joint consideration. In our experiment, we summary the mAP scores of two comparison methods for different cross-modal retrieval tasks: 

$\bullet$ Retrieving audio samples using visual queries (Visual2Audio).

$\bullet$ Retrieving visual samples using audio queries (Audio2Visual).
\subsubsection{Comparison with Existing Approaches}
\begin{figure}[!h]
\centering
\subfloat[]{\label{fig:a}\includegraphics[width = 0.45\textwidth]{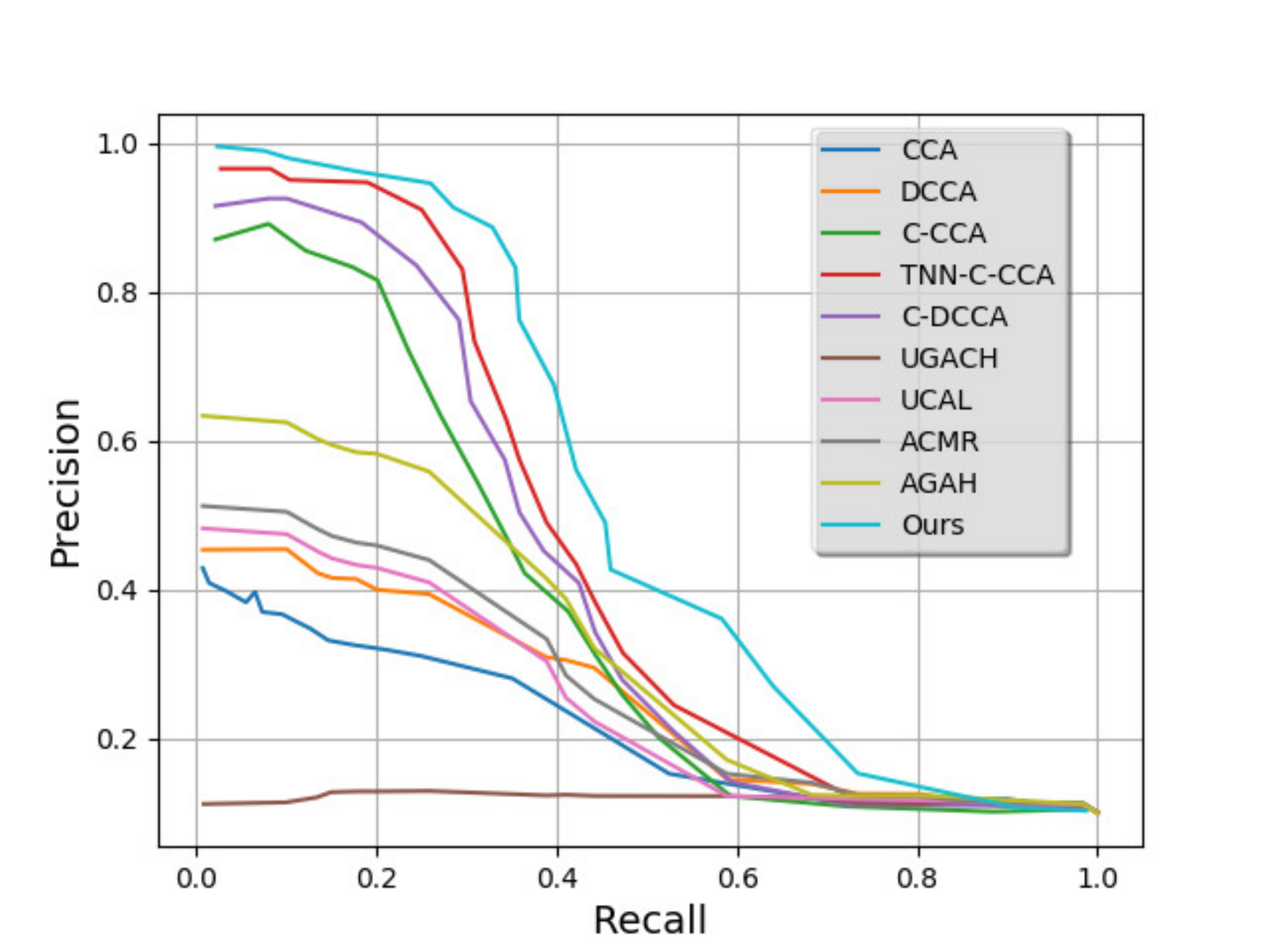}}\quad
\subfloat[]{\label{fig:b}\includegraphics[width = 0.45\textwidth]{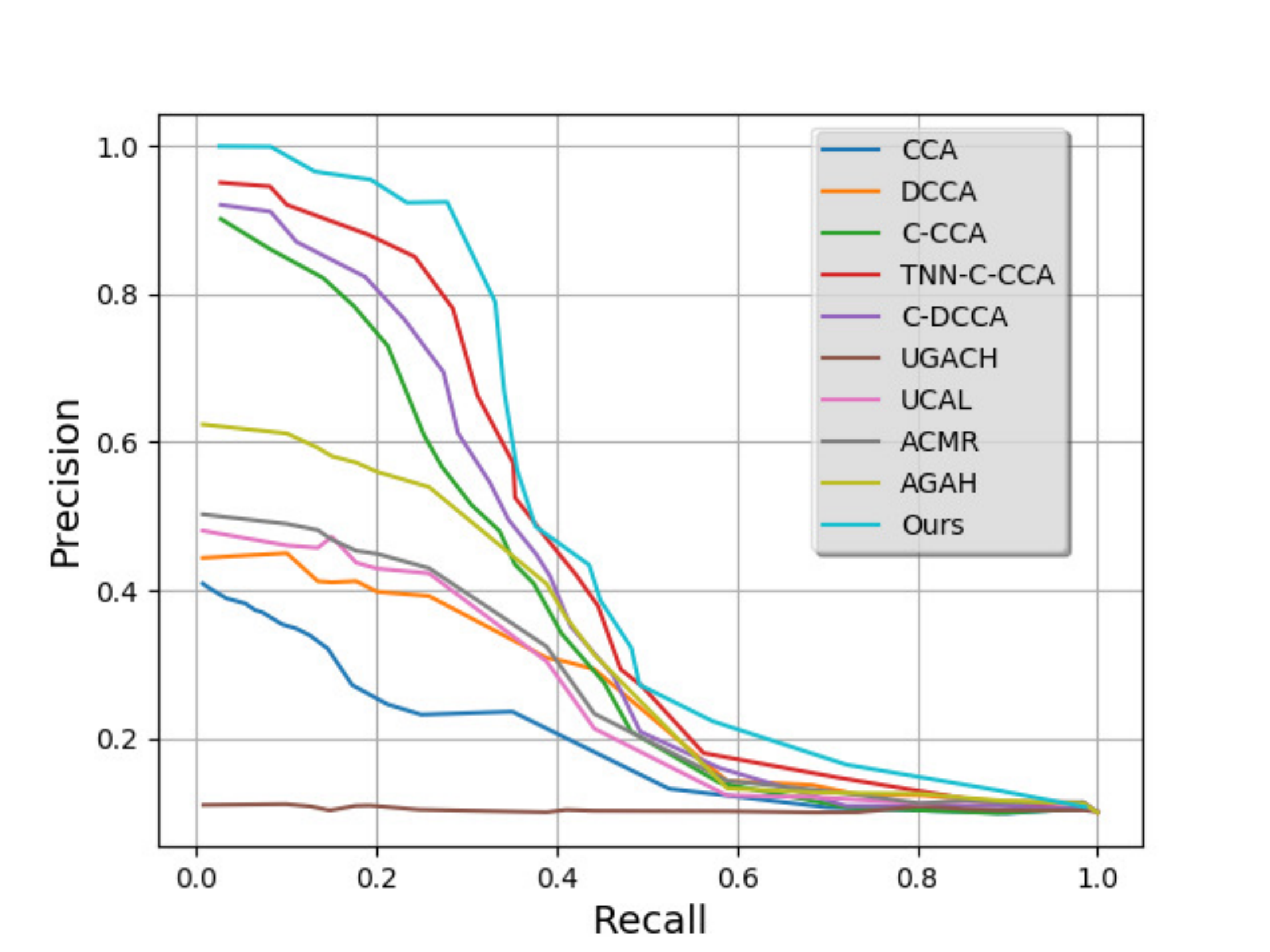}}\\	
\caption{We visualize the experimental values of PRC and compare the other nine different models on the VEGAS dataset. (a) is for audio2visual retrievall and (b) is for visual2audio retrieval.}
\label{fig:retrieval}
\end{figure}

As shown in Table \ref{tab:result1} and Fig. \ref{fig:retrieval}, we utilized the mAP metric and the PRC metric to record the experimental results of the audio-visual cross-modal retrieval on the VEGAS dataset, and visualized the experimental results. To confirm the superiority of our proposed architecture, we compare with eleven existing most advanced cross-modal retrieval methods, including three traditional methods: CCA \cite{hotelling1992relations}, KCCA \cite{lai2000kernel}, and C-CCA \cite{rasiwasia2014cluster}, as well as eight deep learning-basedmethods, namely DCCA \cite{andrew2013deep}, C-DCCA \cite{yu2018category,zeng2018audio} UGACH \cite{zhang2018unsupervised}, AGAH \cite{gu2019adversary}, UCAL \cite{he2017unsupervised}, ACMR \cite{wang2017adversarial}, DSCMR \cite{zhen2019deep} and TNN-C-CCA \cite{zeng2020deep}.
\begin{table}[!ht]
\centering
\caption{Comparison with Existing Approaches on VEGAS dataset in terms of mAP. The highest score is shown in boldface.}
% \resizebox{\linewidth}{!}
\begin{tabular}{c|c c c }\hline
Method         & Audio2Visual  & Visual2Audio & Average   \\ \hline
CCA            & 0.332         & 0.327        & 0.330    \\ 
KCCA           & 0.288         & 0.273        & 0.281    \\  
DCCA           & 0.478         & 0.457        & 0.468    \\ 
C-CCA          & 0.711         & 0.707        & 0.709    \\ 
C-DCCA         & 0.722         & 0.716        & 0.719    \\  \hline
UGACH          & 0.182         & 0.179        & 0.181    \\  
AGAH           & 0.578         & 0.568        & 0.573    \\  
UCAL           & 0.446         & 0.436        & 0.441      \\ 
ACMR           & 0.465         & 0.442        & 0.454      \\ 
DSCMR          & 0.732         & 0.721        & 0.727      \\  
TNN-C-CCA      & 0.751         & 0.738        & 0.745      \\  
\textbf{Ours}  & \textbf{0.811} & \textbf{0.813} & \textbf{0.812}     \\ \hline
\end{tabular} %}
\label{tab:result1}
\end{table}

\begin{table}[!ht]
\centering
\caption{Comparison with Existing Approaches on AVE dataset in terms of mAP. The highest score is shown in boldface.}
% \resizebox{\linewidth}{!}
\begin{tabular}{c|c c c }\hline
Method         & Audio2Visual  & Visual2Audio & Average   \\ \hline
CCA            & 0.190        & 0.189        & 0.190    \\ 
KCCA           & 0.133        & 0.135        & 0.134    \\  
DCCA           & 0.221        & 0.223        & 0.222    \\ 
C-CCA          & 0.153        & 0.152        & 0.153    \\ 
C-DCCA         & 0.230        & 0.227        & 0.229    \\  \hline
UGACH          & 0.165        & 0.159        & 0.162    \\  
AGAH           & 0.200        & 0.196        & 0.198    \\  
UCAL           & 0.153        & 0.150       & 0.152      \\ 
ACMR           & 0.162        & 0.159       & 0.161      \\ 
DSCMR          & 0.314        & 0.256       & 0.285      \\  
TNN-C-CCA      & 0.253        & 0.258       & 0.256      \\  
\textbf{Ours}  & \textbf{0.358} & \textbf{0.343}& \textbf{0.350}     \\ \hline
\end{tabular} %}
\label{tab:result2}
\end{table}
$\bullet CCA$ \cite{hotelling1992relations} projects the features of different modalities into a common subspace, and realizes cross-modal retrieval tasks by maximizing the correlation between modal samples.

$\bullet KCCA$ \cite{lai2000kernel} improves the CCA algorithm by introducing the concept of a "kernel trick" for learning common space representation. We utilize Gaussian kernel as the kernel function of CCA in the comparison experiment.

$\bullet C-CCA$ \cite{rasiwasia2014cluster} (Cluster-CCA) clusters different modalities data into several categories, and try to enhance the correlation intra-cluster.

$\bullet DCCA$ \cite{andrew2013deep} utilizes deep learning network to solve nonlinear projection problem, apply CCA-like objective function to maximize the correlation of different modal samples.

$\bullet C-DCCA$ in \cite{yu2018category,zeng2018audio} is a combination of Cluster-CCA and DCCA, which learn the non-linear representation by deep learning method into several related clusters to optimize the correlation.

$\bullet UGACH$ \cite{zhang2018unsupervised} utilizes GAN network to extract potential features of cross-modal data.

$\bullet AGAH$ \cite{gu2019adversary} the multi-label attention module of adversarial learning applies to enhance the ability to distinguish between cross-modal representations.

$\bullet UCAL$ \cite{he2017unsupervised} maximizes the correlation between visual-text modalities in the common representation space. The classifier predicts the visual-text modalities of the learned features, and applies adversarial learning ideas to add regularization to the model.

$\bullet ACMR$ \cite{wang2017adversarial} adopts adversarial training ideas, utilizing classifiers and feature projections to guide the model to learn modal invariant and discriminative features.

$\bullet DSCMR$ \cite{zhen2019deep} guides the network to learn the discriminative features in the category subspace and feature common subspace through the method of supervised learning.

$\bullet TNN-C-CCA$ \cite{zeng2020deep} uses the triple loss function to reduce the category spacing of features in the common representation subspace, thereby improving the retrieval performance of Cluster-CCA.

The performances of our proposed method and comparison algorithms on AVE datasets are reported in Table \ref{tab:result2}. From the presented results, we can have the following observations: 

$\bullet$ Our proposed architecture can achieve the best performance on the AVE datasets in terms of mAP, which verifies the impact of improved retrieval performance via the correlations between different modalities data. 

$\bullet$ Our proposed architecture is extensively better than both CCA \cite{hotelling1992relations} and KCCA \cite{lai2000kernel} among most cases, e.g., on the AVE dataset, CCA and KCCA are only 0.190 and 0.134 for mAP. 
That is because they are unsupervised learning methods, which do not consider discriminative label information of different modalities. 

$\bullet$ The reason for the poor clustering overall performance of DCCA is that it can not reconstruct the information to make sure that the representation after encoding the network can nevertheless reflect the structure of the original data, which causes mAP to be 0.222 on the AVE dataset.
Additionally, our proposed architecture also performs better than DSCMR and TNN-C-CCA because these two methods cannot make full use of the correlations among the inter-modal data.

\subsubsection{Convergence Analysis} 
\begin{figure}[ht]
\centering
\subfloat[VEGAS dataset]{\label{fig:a}\includegraphics[width = 0.4\textwidth]{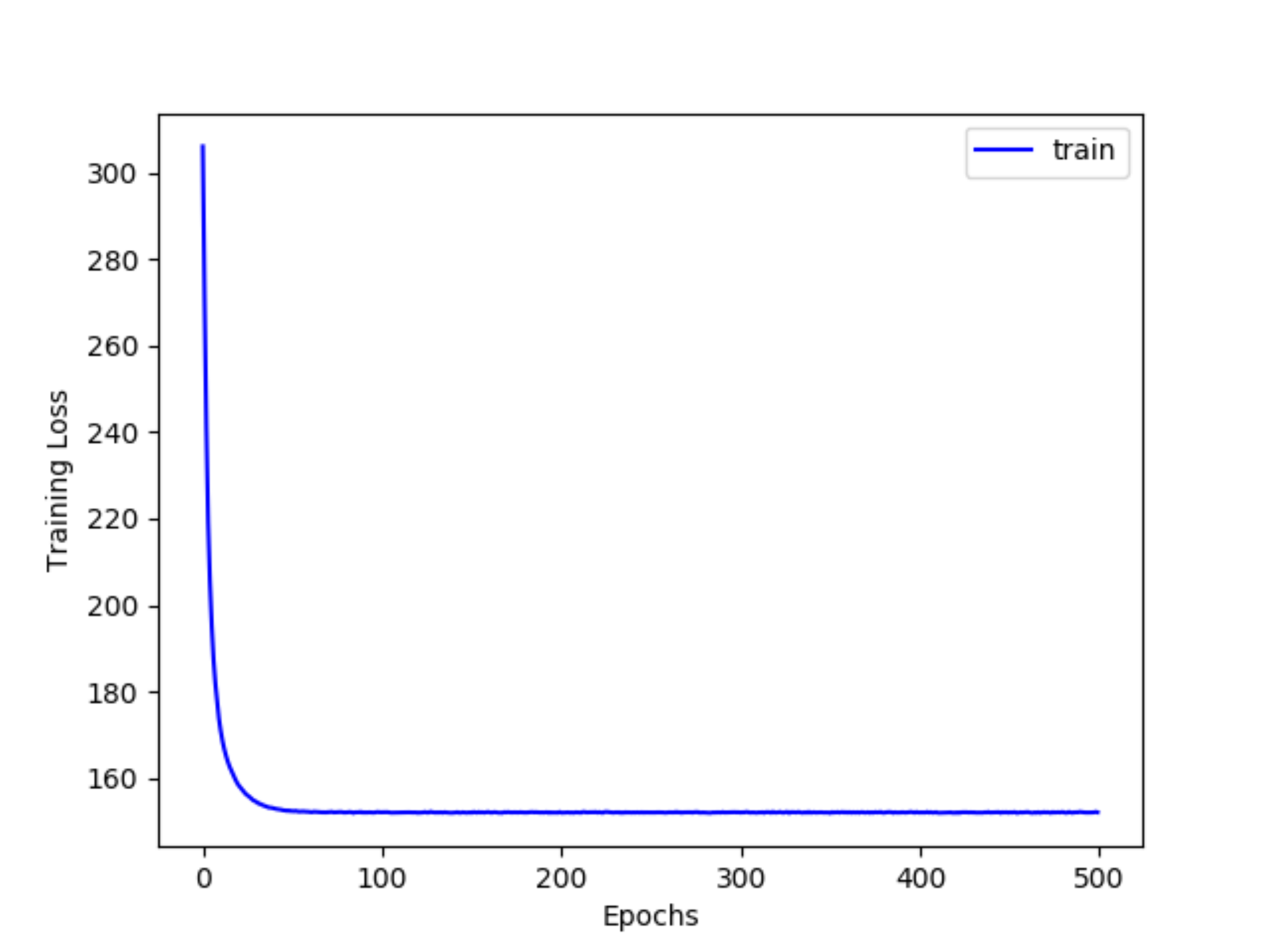}}\quad
\subfloat[AVE dataset]{\label{fig:b}\includegraphics[width = 0.4\textwidth]{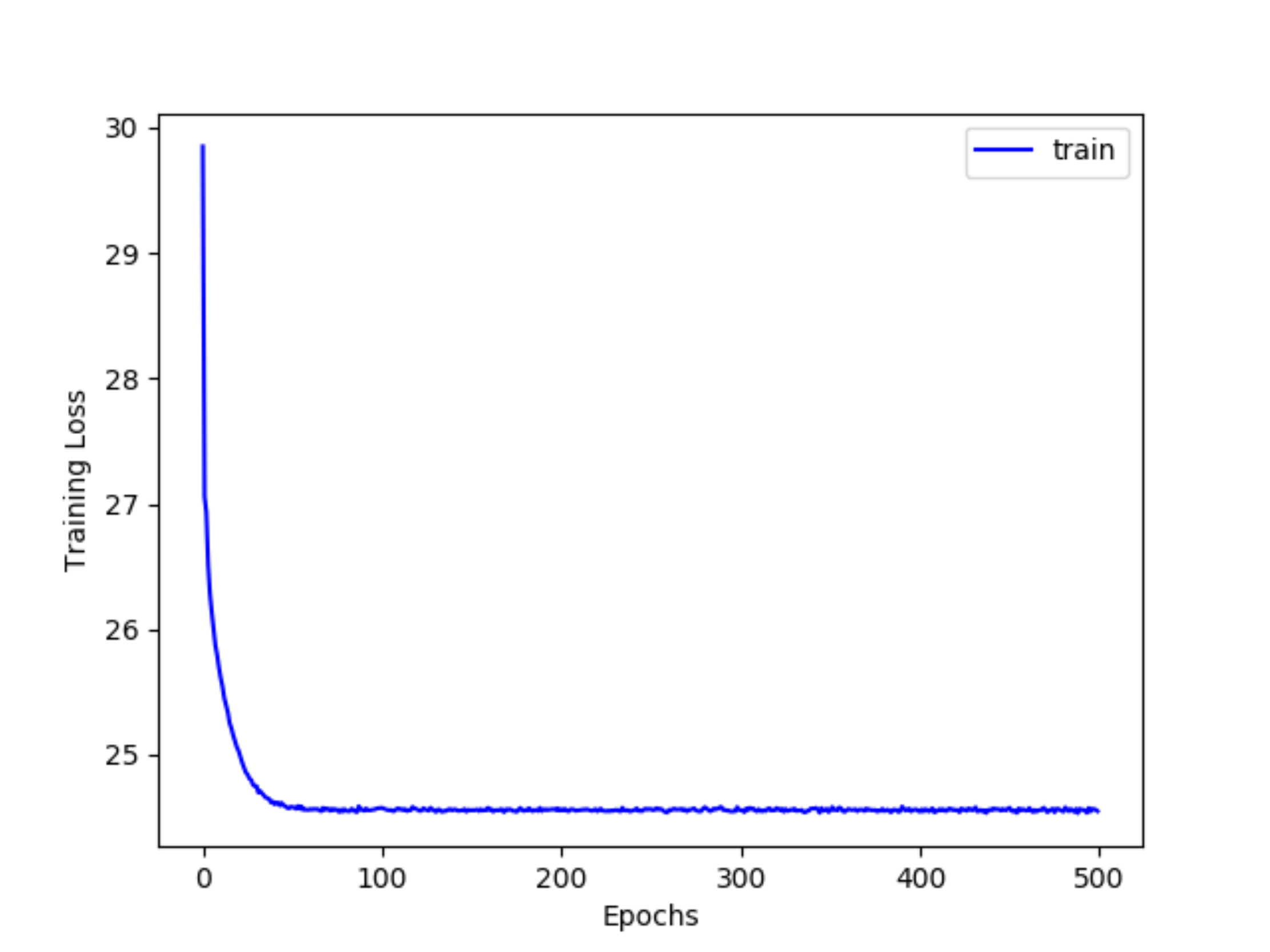}}	
\caption{We visualize the loss curve of our architecture on the VEGAS and AVE datasets. And set 500 epochs to train the entire network and obtain a loss value for each 100 epochs.}
\label{fig:VEGASdataset}
\end{figure}

In this section, we train the proposed architecture on the VEGAS and AVE datasets, and use the output loss value to draw the change curve of the total loss value of the objective function to investigate the convergence of the proposed architecture.
As shown in Fig. \ref{fig:VEGASdataset}, the value of the total loss function decreases as the number of iterations increases, and the values approach to be a fixed value after a few iterations (less than 20 iterations),
where each iteration includes 100 epochs. Therefore, our proposed optimization algorithm is reliable and converges quickly.

\subsubsection{Parameters Analysis}
In our architecture, there are four regularization parameters $\lambda_{1}$, $\lambda_{2}$, $\lambda_{3}$ and $\lambda_{4}$, we simultaneously adjust them to obtain the best model, However, for simplicity and evaluating the effect of each parameter in our experiments, we fix one and vary the other three for each time. 

\begin{figure}[!h]
\centering
\subfloat[]{\label{fig:a}\includegraphics[width = 0.4\textwidth]{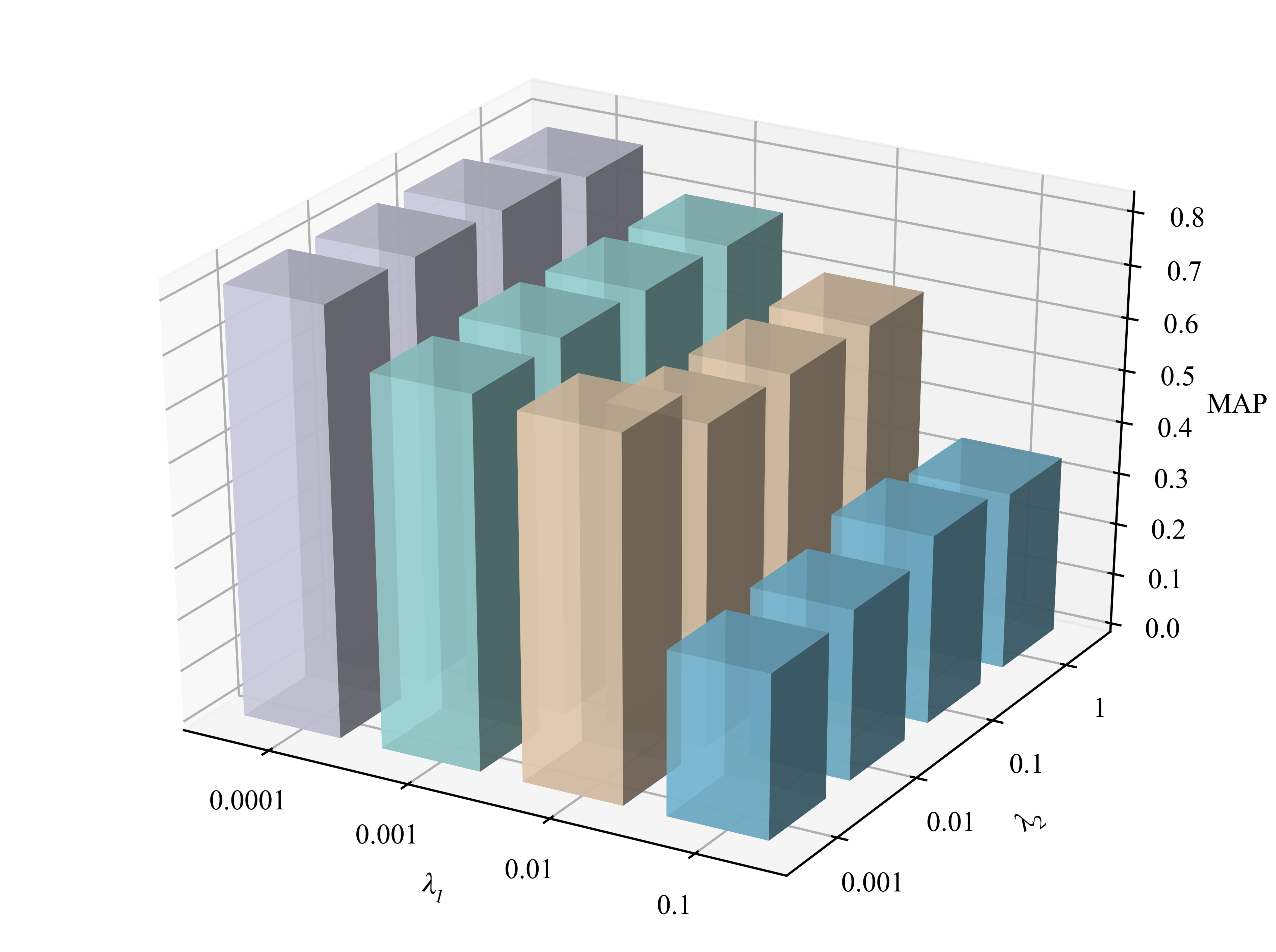}}\quad
\subfloat[]{\label{fig:b}\includegraphics[width = 0.4\textwidth]{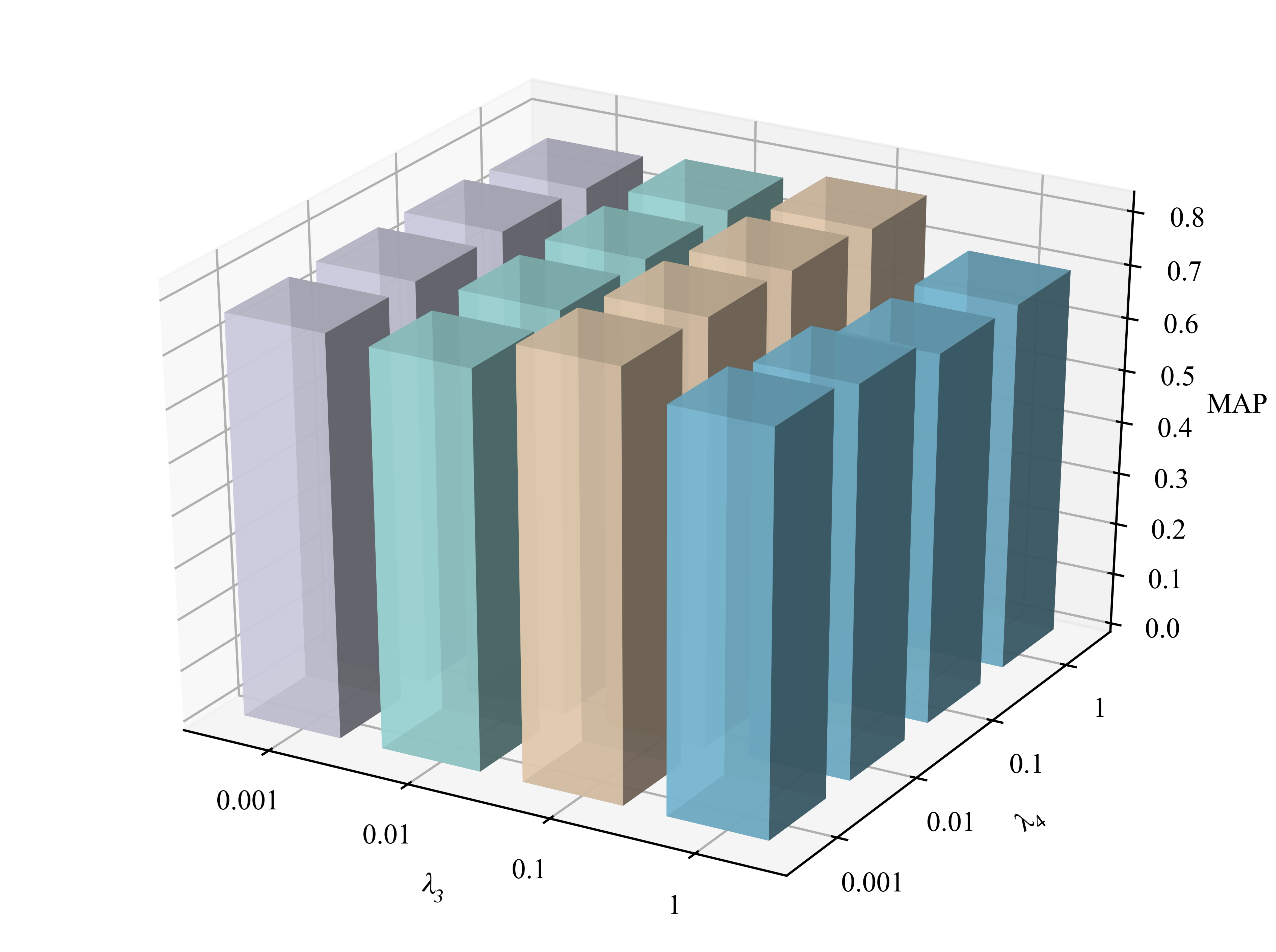}}\\	
\caption{The effect of parameters $\lambda_{1}$, $\lambda_{2}$, $\lambda_{3}$ and $\lambda_{4}$ on VEGAS dataset, where $\lambda_{1}$ is the regularization parameter of the VAE calculation, $\lambda_{2}$ is the regularization parameter of the correlation calculation, $\lambda_{3}$ is the regularization parameter of the latent loss error and $\lambda_{4}$ is the regularization parameter of the center calculation. (a) is the retrieval results in terms of mAP, when fixing $\lambda_{3}$ , $\lambda_{4}$  and varying  $\lambda_{1}$ and $\lambda_{2}$. (b) is retrieval results in terms of mAP, when fixing $\lambda_{1}$, $\lambda_{2}$, and varying $\lambda_{3}$, and $\lambda_{4}$.}
\label{fig:parameters}
\end{figure}

Firstly, we set the regularization parameters of the VAE calculation $\lambda_{1}$ and modify the regularization parameters of the correlation error, the distance error and the center error $\lambda_{2}$, $\lambda_{3}$ and $\lambda_{4}$ in range $\left\{ 0.001, 0.01, 0.1, 1 \right\}$. Then we fix $\lambda_{2}$ and also vary $\lambda_{3}$ and $\lambda_{4}$ in the same range. Since the approach of adjusting the parameters is the equal on both datasets, we solely exhibit the changes of the parameters in the VEGAS dataset. From the Fig. \ref{fig:parameters}, we notice that:

$\bullet$ Our architecture can obtain the best mAP values on VEGAS dataset when $\lambda_{1}= 0.0001$, $\lambda_{2}=0.001$, $\lambda_{3}=0.1$ and $\lambda_{4}=0.01$;

$\bullet$ In our architecture, when $\lambda_{3}$ and $\lambda_{4}$ are fixed, the change range of $\lambda_{1}$ and $\lambda_{2}$ is relatively large; when $\lambda_{1}$ and $\lambda_{2}$ are fixed, the change range of $\lambda_{3}$ and $\lambda_{4}$ is relatively small. 

\subsubsection{Role of Category}
\begin{figure}[ht]
\centering
\subfloat[]{\label{fig:a}\includegraphics[width = 0.6\textwidth]{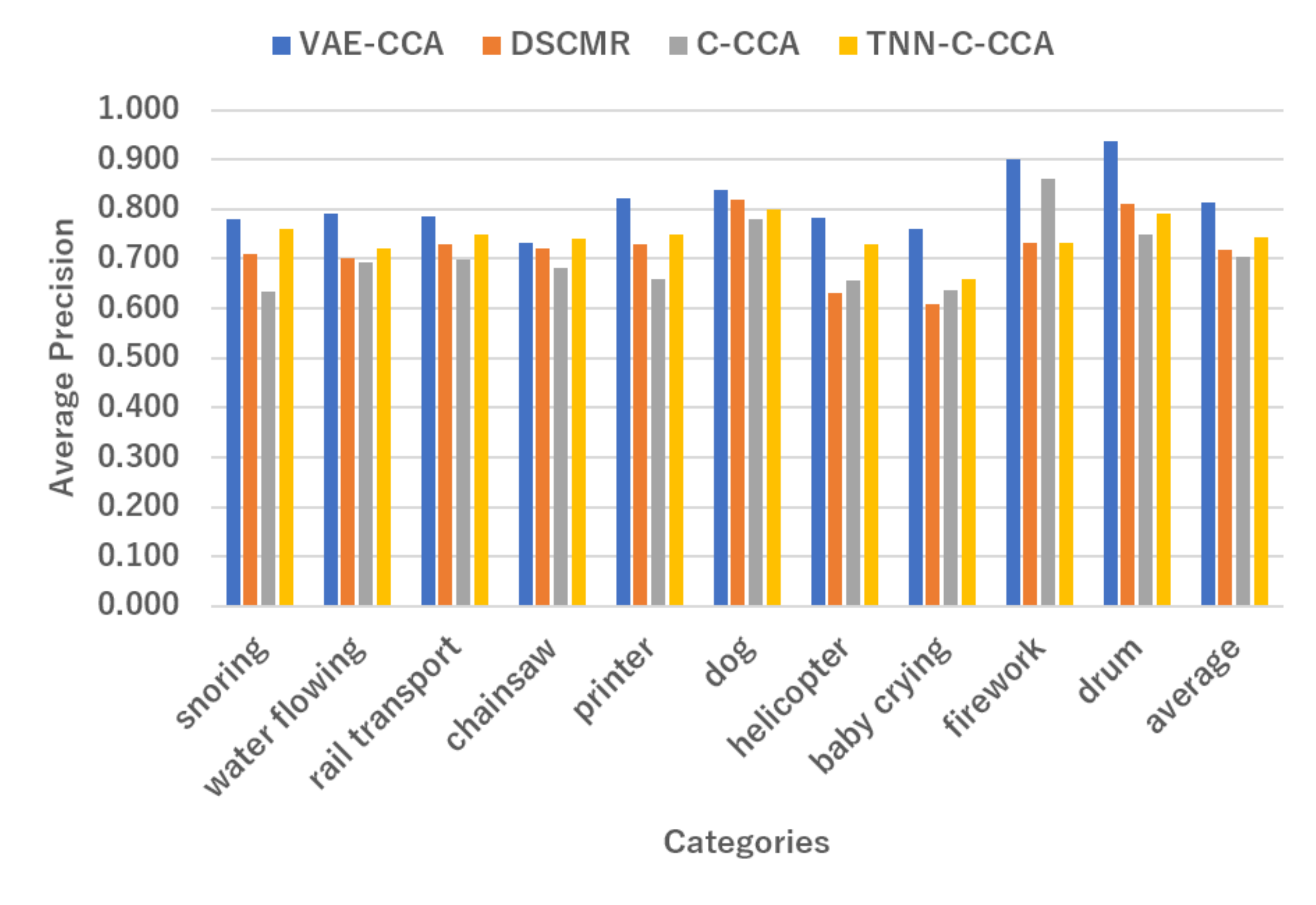}}\\
\subfloat[]{\label{fig:b}\includegraphics[width = 0.6\textwidth]{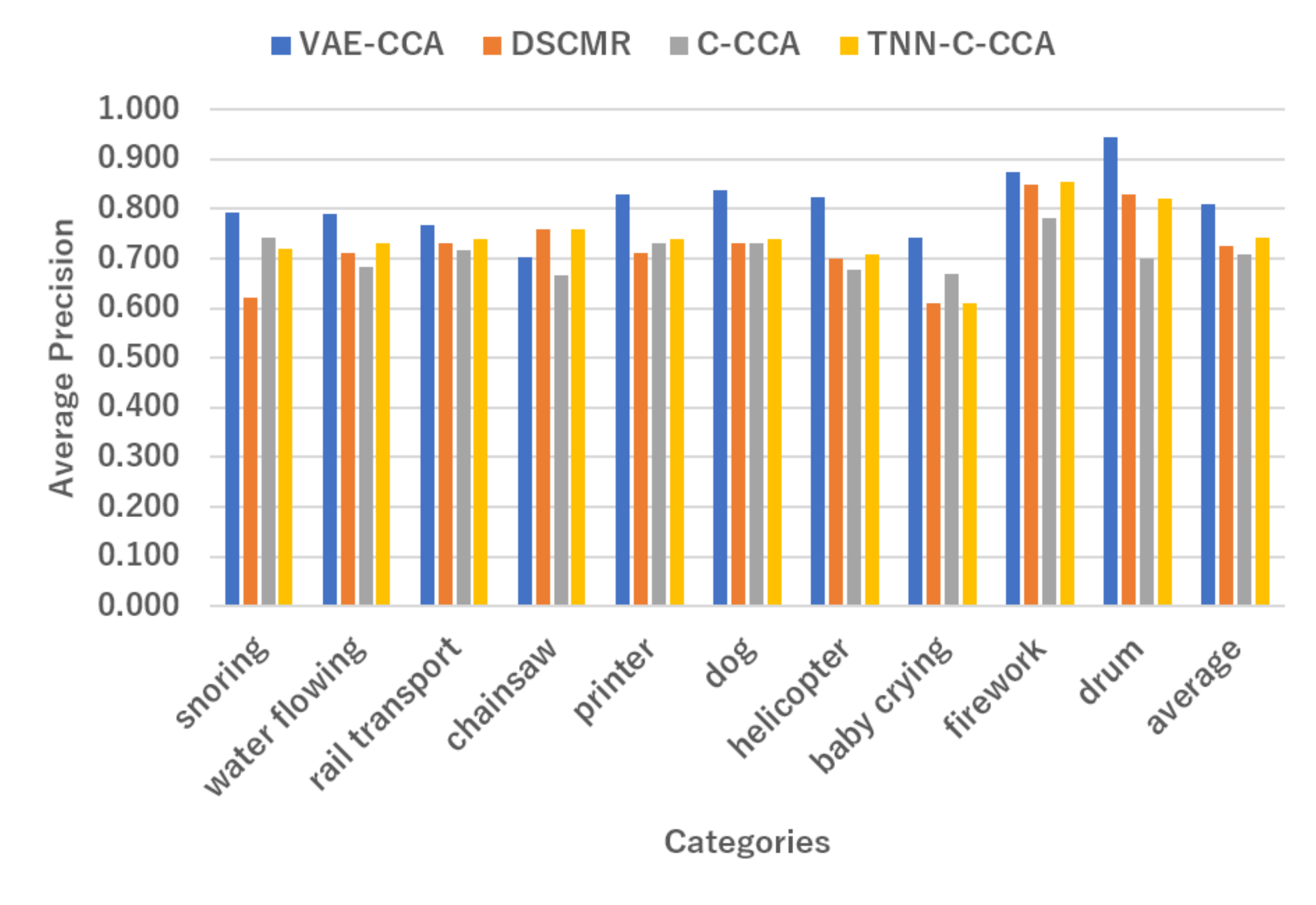}}
\caption{We compare the proposed method with the C-CCA, TNN-C-CCA, and DSCMR methods in the VEGAS dataset, and show the results of each category. According to the visualized statistical results, the retrieval accuracy rates of different categories are quite different. (a) visual2audio and in (b) audio2visual. }
\label{fig:category}
\end{figure}

In addition, we investigate the effectiveness of common subspace representation via audio-visual retrieval tasks. Fig. \ref{fig:category} shows the average accuracy (AP) score of each category search after comparing our model with C-CCA, TNN-C-CCA and DSCMR models on the VEGAS dataset. It can be roughly seen in Fig. \ref{fig:category} that the higher the AP value is, the easier it is to retrieve, and the retrieval accuracy of different categories varies greatly. When the audio is utlized as a query, we can see that ``baby crying'' and ``drum'' use the proposed method to get the highest AP. When using the our proposed architecture (VAE-CCA) for retrieving the ``helicopter'', audio2visual can reach 82.3$\%$, while visual2audio is only 78.3$\%$, a difference of nearly 4.0$\%$. By comparing with methods such as DSCMR, TNN-C-CCA, and C-CCA, our proposed architecture can obtain retrieval accuracy higher than other methods in each category, which verifies the effectiveness of our architecture.

\subsubsection{Ablation Study}
The total loss function of our proposed architecture is composed of multiple loss functions to minimize the discriminative loss in the category latent space and the common representation subspace, and the distance loss and centre loss in the common representation subspace, respectively. We conducted ablation experiments on the VEGAS data set to examine the impact of the different loss functions on the whole performance of the proposed architecture. The experimental results are shown in the Table \ref{tab:result3}.

From Table \ref{tab:result3} we can observe: 

$\bullet$ Correlation constraints have a positive affect on retrieval performance. They maximize the correlation of the information between the modalities and obtain a better representation of the common subspace.

$\bullet$ The distance loss has an impact on on the proposed architecture. Based on the above experimental results, all the loss functions in our proposed architecture can be combined to obtain properly retrieval performance.

\begin{table*}[ht]
\centering
\caption{Ablation study on VEGAS dataset in terms of mAP.}
% \resizebox{\linewidth}{!}
\begin{tabular}{c c c c }\hline
Methods                          & Audio2Visual  & Visual2Audio & Average  \\ \hline
With center loss                 & 0.130         & 0.145        & 0.138   \\ 
With correlation loss            & 0.520         & 0.404        & 0.462    \\ 
With distance                    & 0.651         & 0.638        & 0.644    \\  
Full our proposed method         & 0.811         & 0.813        & 0.812    \\ \hline
\end{tabular} %}
\label{tab:result3}
\end{table*}

\subsubsection{Visualization of the Learned Representation}
\begin{figure}[ht]
\centering 
\includegraphics[width=.8\textwidth,height=0.6\textwidth]{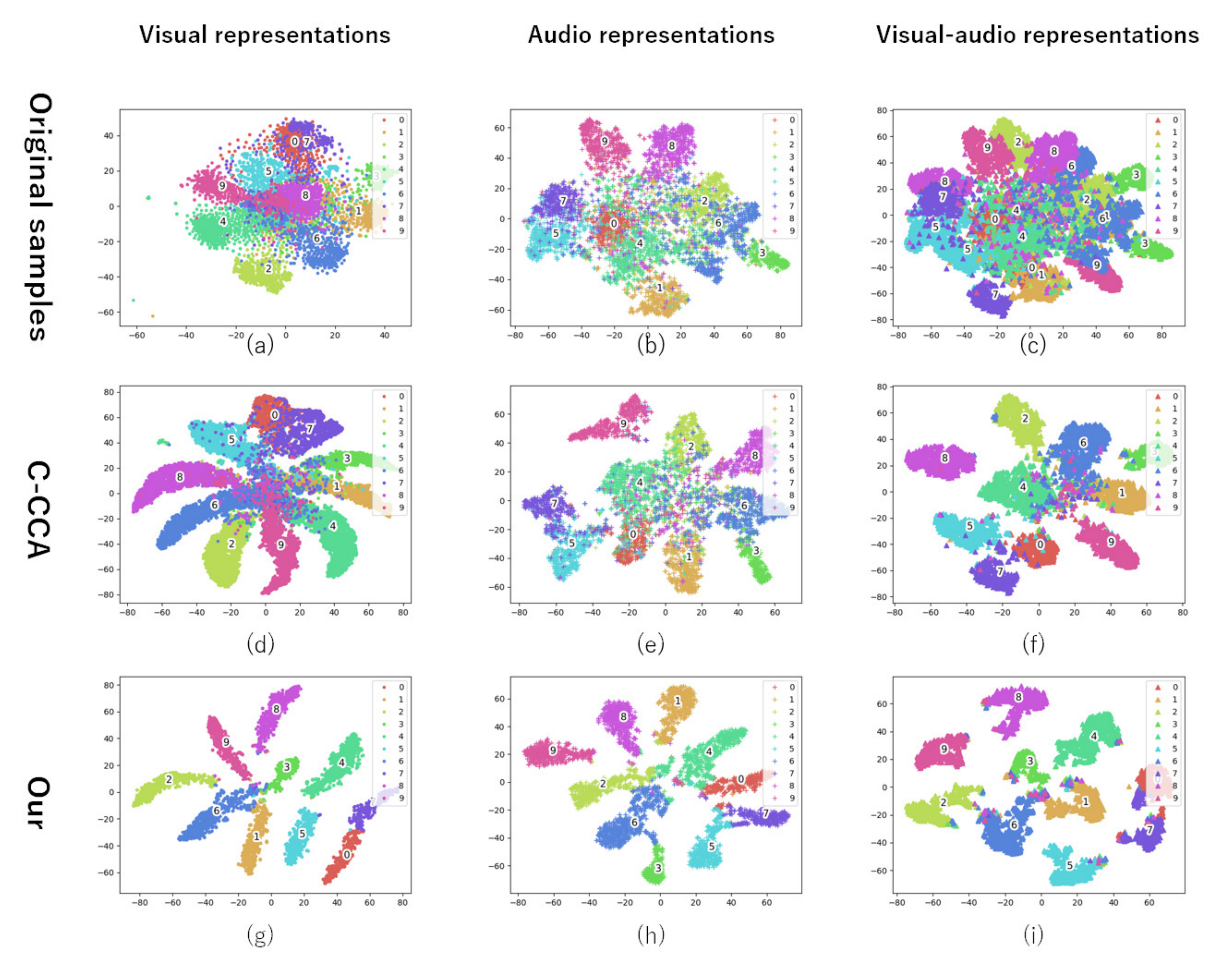}
\caption{The t-SNE method is used to visualize the audio-visual test data in the VEGAS dataset. The circles represent samples from visual and audio modality. We utilize the same color to represent samples belonging to the same semantic category, visualize the learned representation subspace via utilizing the t-SNE approach, showing the original data distribution of audio, visual, and audio-visual. audio-visual, and visualize the feature representaton subspace learned by the C-CCA method and the feature representation subspace learned by our proposed method. The circle represents audio data, and the cross represents visual data, the triangle sign represents visual-audio data.}
\label{fig:visual}
\end{figure}

Here we adopt the t-SNE \cite{van2008visualizing} approach to map the common representation subspace of audio and visual samples into a two-dimensional visualization plane to learn about the effectiveness of the proposed architecture. Fig. \ref{fig:visual}(a)-(c) exhibit the distribution of original 1,024-dimensional visual features and 128-dimensional audio features. We can see that the distribution of visual modalities and audio modalities in the VEGAS Dataset is largely different, and the sample category spacing is small. It is not always easy to distinguish the sample categories effectively.
%To visually investigate the effectiveness of the proposed method, we adopt the t-SNE \cite{van2008visualizing} approach to embed the representations of the visual and audio samples (in the common representation subspace) into a two-dimensional visualization plane. The results of the original 1,024-dimensional features visual and 128-dimensional audio features (after the embedding process) are displayed in Fig. \ref{fig:visual}(a), Fig. \ref{fig:visual}(b) and Fig. \ref{fig:visual}(c) respectively. We can see that the distributions of the visual modality and the audio modality in the VEGAS Dataset are largely different, and the samples are hard to be clustered in the original input space.
Fig. \ref{fig:visual}(d)-(f) show the two-dimensional distributions of the visual and audio representations in the common subspace by C-CCA method. from Fig. \ref{fig:visual}(e)-(f) We can see that C-CCA embeds category information into the feature representation space, where clusters of different colors represent different categories. It can be observed that these clusters are not completely distinguishable. From the visualization results, the embedding of our VAE-CCA is much better than C-CCA embedding. The function of distinguishing loss in common subspace and semantic category subspace can distinguish samples from different semantic categories, and efficaciously divide the samples of joint space into independent semantic groups.

%In addition, we studied the effectiveness of public space expression through audiovisual retrieval tasks. We try to compare the retrieval results of our model with the other three best models. Fig. \ref{fig:visuals} provides examples of audio-to-visual retrieval generated by ACMR, AGAH, TNN-C-CCA and our model for a given audio on the VEGAS dataset, with the "chain saw" tag as the query. We can observe that the AP of our model is 87.1\% in all ranking lists. For other models, the AP of the ACMR model is 41.6\% in all ranking lists; the AP of the AGAH model is 56.3\% in all ranking lists.

\begin{figure}[htbp]
\centering 
\includegraphics[width=.8\textwidth,height=0.6\textwidth]{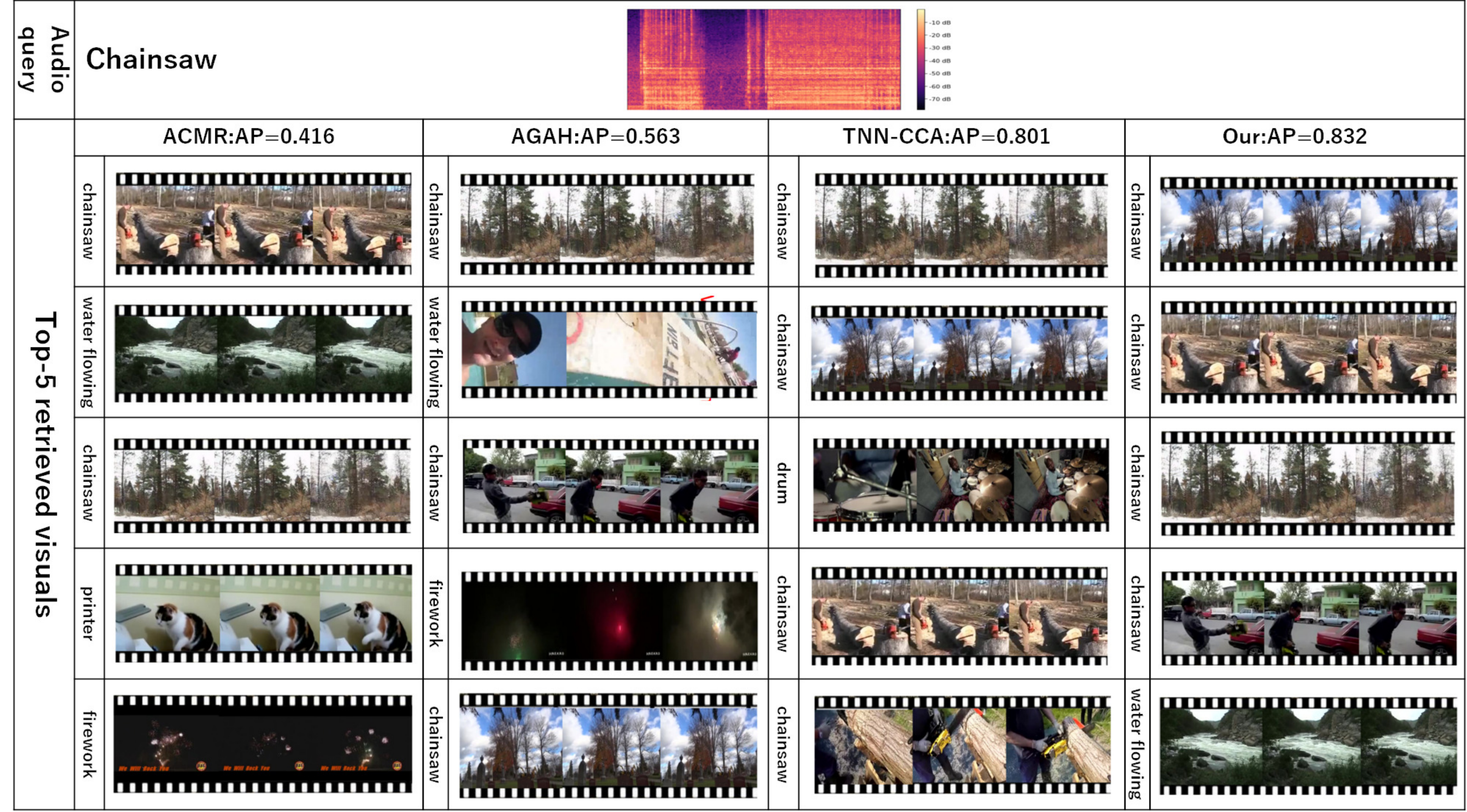}
\caption{Visualize the results of audio2visual retrieval: compare our proposed architecture with the other three best existing methods TNN-CCA, AGAH and ACMR models. Apply audio as a query and display the top five retrieved visuals.}
\label{fig:visuals}
\end{figure}

In addition, to verify the effectiveness of common subspace representation, we conducted audio2visual retrieval experiments. Comparing our model with the other three best models on the VEGAS dataset, Fig. \ref{fig:visuals} shows the audio2visual retrieve results of ACMR, AGAH, TNN-C-CCA, and our model. We apply the audio ``chainsaw'' as the query, and we can observe that the AP of our model is 83.2\% of all ranking lists. For other models, the AP of the ACMR model is 41.6\% of all rankings; the AP of the AGAH model is 56.3\% of all rankings.

\begin{figure}[!ht]
\centering
\subfloat[]{\label{fig:a}\includegraphics[width=1.8in]{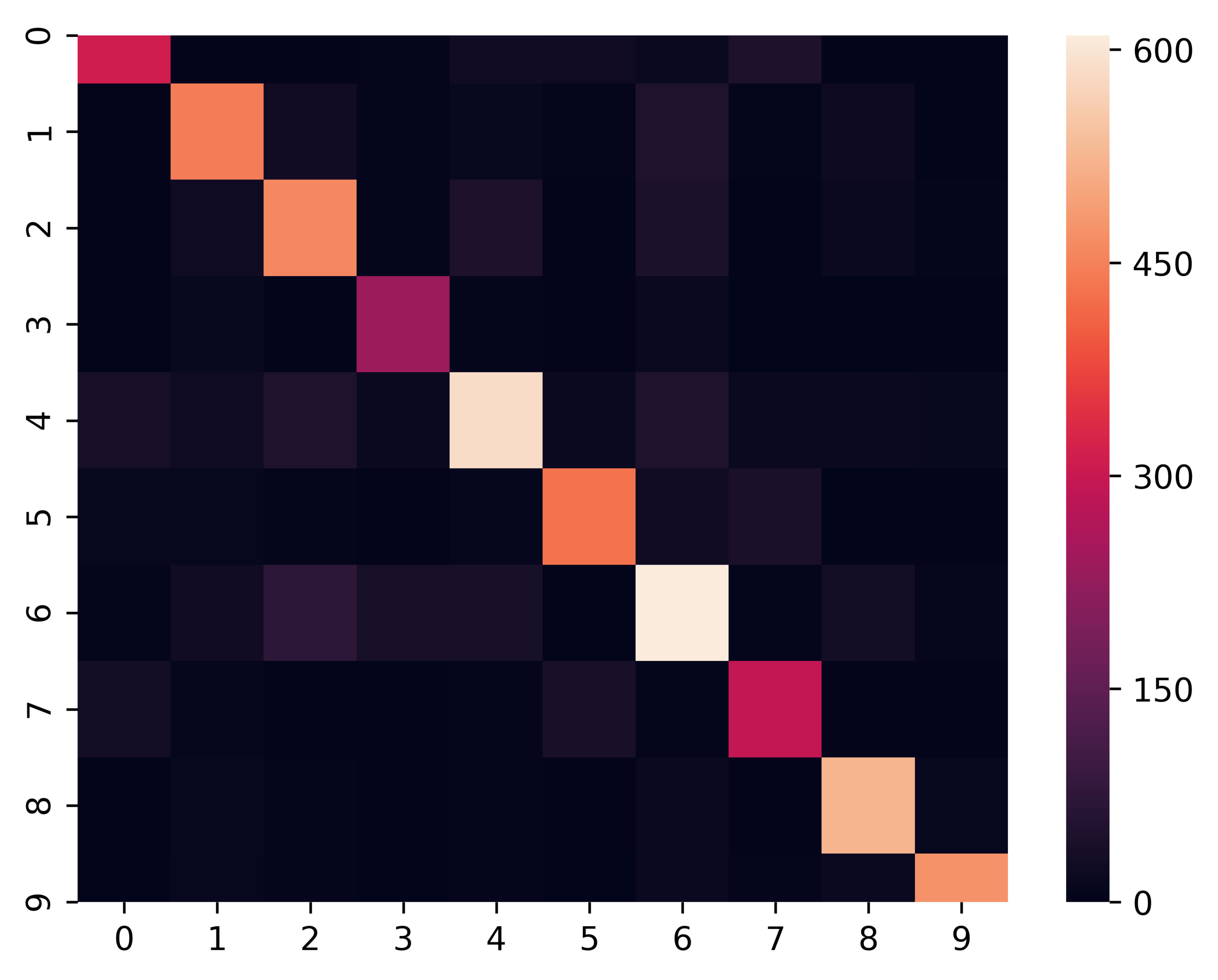}}
\subfloat[]{\label{fig:b}\includegraphics[width=1.8in]{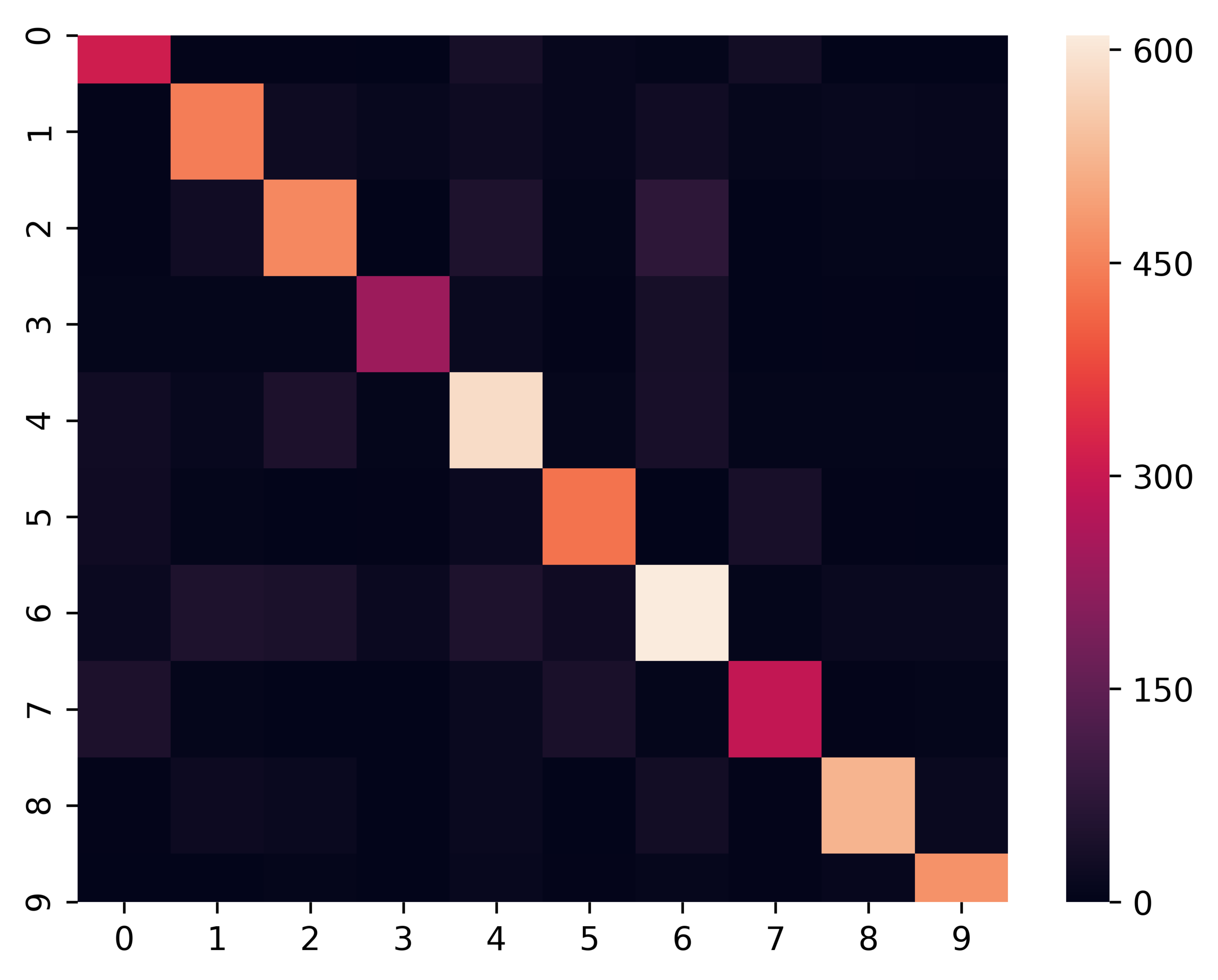}}	
\caption{ The confusion matrix achieved on the VEGAS dataset with our proposed architecture. The figure (a) is for audio2visual retrieval, the figure (b) is for visual2audio retrieval.}
\label{fig:matrix}
\end{figure}

Finally, we adopt the confusion matrix evaluation metric on the VEGAS dataset to conduct video and audio cross-modal retrieval experiments to evaluate the accuracy of the proposed cross-modal retrieval architecture and then calculate the confusion matrix and visualized the results. As shown in Fig. \ref{fig:matrix}, the number of retrieved samples is proportional to the color brightness. The horizontal axis represents the predicted category, and the vertical axis represents the actual category. The number of samples retrieved correctly is displayed on the diagonal, and the number retrieved incorrectly is displayed on the area outside the diagonal. From our experimental results, the correctly retrieved samples are concentrated on the diagonal, which verifies that our proposed architecture has better retrieval performance.

%\begin{figure*}[ht]
%	\centering 
%	\includegraphics[width=.5\textwidth]{images/confusion}
%	\caption{The confusion matrix achieved on the VEGAS dataset with our proposed architecture. The figure (a) is for audio2visual retrieval, the figure (b) is for visual2audio retrieval.}
%	\label{fig:matrix}
%\end{figure*}
\section{Conclusion}
In this paper, we propose a novel architecture based on VAE (VAE-CCA) to guide the model to learn more discriminative features via making full use of semantic category information and modal-invariant representations of different modalities data. We apply CCA to learn audio-visual correlation embedding in the mutual latent space. Our architecture can mitigate the discrepancy between audio-visual data while capturing discriminative features, and making sure that the representation processed by way of the encoder will keep the traits of the unique data. Since the wide range of datasets, there are problems such as noise and lack of information, which lead to uncertainty in the data. In this paper, we utilize probabilistic modeling methods to deal with this issue. We carried out many comparative experiments on two benchmark datasets and comprehensively analyzed the experimental results to exhibit the effectiveness of the proposed cross-modal retrieval architecture.

In future research, we desire to apply our architecture to different kinds of cross-mode retrieval, such as visual2text, audio2text, and video2text. We will additionally consider to apply adversarial learning methods to enhance our cross-modal retrieval performance, and strive to extend our current architecture to achieve cross-modal data fusion to address the issue of insufficient data.

\bibliographystyle{IEEEtran}
\bibliography{IEEEabrv,reference}

\end{document}